\documentclass[aps,twocolumn,groupedaddress,showpacs]{revtex4}
\usepackage{graphicx}
\begin{document}
%%%%%%%%%%%%%%%%%%%%%%%%%%%%%%%%%%%%%%%%%%%%%%%%%%%%%%%%%%%%%%%%%%%%%%%%%%%%%
% You should use BibTeX and revtex.bst for references
%\bibliographystyle{apsrev}
%%%%%%%%%%%%%%%%%%%%%%%%%%%%%%%%%%%%%%%%%%%%%%%%%%%%%%%%%%%%%%%%%%%%%%%%%%%%%
% marks overfull lines with blackboxes
%\draft - no longer supported, use the 'draft' option instead

%\preprint{}
\title{Incommensurate state in a quasi-one-dimensional $S=1/2$ bond-alternating antiferromagnet with frustration in magnetic fields}% Optional argument for running titles on pages
%\title[]{}

\author{Takahumi Suzuki and Sei-ichiro Suga}
%\thanks{}
\affiliation{Department of Applied Physics, Osaka University, Suita, Osaka 565-0871, Japan}
\date{\today}
%%%%%%%%%%%%%%%%%%%%%%%%%%%%%%%%%%%%%%%%%%%%%%%%%%%%%%%%%%%%%%%%%%%%%%%%%%%%%
%                         ABSTRACT                                          %
%%%%%%%%%%%%%%%%%%%%%%%%%%%%%%%%%%%%%%%%%%%%%%%%%%%%%%%%%%%%%%%%%%%%%%%%%%%%%
\begin{abstract}
We investigate the critical properties of the $S=1/2$ bond-alternating spin chain with a next-nearest-neighbor interaction in magnetic fields. By the numerical calculation and the exact solution based on the effective Hamiltonian, we show that there is a parameter region where the longitudinal incommensurate spin correlation becomes dominant around the half-magnetization of the saturation.  
Possible interpretations of our results are discussed. 
We next investigate the effects of the interchain interaction ($J^{\prime}$). The staggered susceptibility and the uniform magnetization are calculated by combining the density-matrix renormalization group method with the interchain mean-field theory. 
For the parameters where the dominant longitudinal incommensurate spin correlation appears in the case $J^{\prime}=0$, the staggered long-range order does not emerge up to a certain critical value of $J^{\prime}$ around the half-magnetization of the saturation. 
We calculate the static structure factor in such a parameter region. 
The size dependence of the static structure factor at $k=2k_{\rm F}$ implies that the system has a tendency to form an incommensurate long-range order around the half-magnetization of the saturation. 
We discuss the recent experimental results for the NMR relaxation rate in magnetic fields performed for pentafluorophenyl nitronyl nitroxide. 
\end{abstract}
%%%%%%%%%%%%%%%%%%%%%%%%%%%%%%%%%%%%%%%%%%%%%%%%%%%%%%%%%%%%%%%%%%%%%%%%%%%%%
%\maketitle must follow title, authors, abstract and \pacs
\maketitle
%\section{}
%\label{}
%\subsection{}
%\subsubsection{}
%%%%%%%%%%%%%%%%%%%%%%%%%%%%%%%%%%%%%%%%%%%%%%%%%%%%%%%%%%%%%%%%%%%%%%%%%%%%%
%                        MAIN TEXT                                          %
%%%%%%%%%%%%%%%%%%%%%%%%%%%%%%%%%%%%%%%%%%%%%%%%%%%%%%%%%%%%%%%%%%%%%%%%%%%%%
%

%%%%%%%%%%%%%%%%%%%%%%%%%%%%%%%%%%%%%%%%%%%%%%%%%%%%%%%%%%%%%%%%%%%%%%%%%
\section{INTRODUCTION}
%%%%%%%%%%%%%%%%%%%%%%%%%%%%%%%%%%%%%%%%%%%%%%%%%%%%%%%%%%%%%%%%%%%%%%%%%
One-dimensional (1D) spin-gapped systems have attracted a great amount of attention both theoretically and experimentally. 
It was shown for the 1D $S=1/2$ spin-gapped systems that a noticeable feature of each system appears in critical properties, when the energy gap is collapsed by external magnetic fields \cite{CG,ES}. 
In $H_{c_1} <H < H_{c_2}$, where $H_{c_1}$ and $H_{c_2}$ are the lower critical field and the saturation field respectively, the field dependence of the critical exponent of the spin correlation function was calculated numerically for the Haldane-gap system \cite{ST1,KF}, the $S=1/2$ bond-alternating chain \cite{Sakai1}, the alternating-spin chain \cite{kura}, and the $S=1/2$ two-leg ladder \cite{US2,HF}. 
It was shown that in these systems the transverse staggered spin correlation makes a leading contribution, and that the critical exponent exhibits characteristic behavior as function of magnetic field or the magnetization in each system. 
Such features can be observed by the field dependence of the divergence exponent of the NMR relaxation rate $1/T_1$ with decreasing temperature \cite{CG,HS}. Experimentally, the divergence exponent of $1/T_1$ was measured in the $S=1/2$ two-leg ladder \cite{Chab} and the Haldane-gap system \cite{Goto}. 
The results were discussed in connection with the theoretical results \cite{CG,ST1,US2,GT,HS}.

When temperature is further decreased and the interchain or interladder interaction becomes relevant, a quantum phase transition towards a three-dimensional (3D) ordered state takes place. 
Such 3D ordered states in magnetic fields were observed experimentally in several quasi-1D spin-gapped systems such as Haldane-gap materials NDMAZ \cite{NDMAZ} and NDMAP \cite{NDMAP}, a $S=1/2$ two-leg spin-ladder material CuHpCl \cite{Hamm}, and $S=1/2$ bond-alternating chains ${\rm Cu(NO_3)_2 \cdot 2.5H_2O}$ \cite{AFAF} and ${\rm (CH_3)_2 CHNH_3 CuCl_3}$ \cite{FAF}.

The $S=1/2$ bond-alternating spin chain with a next-nearest-neighbor (NNN) interaction is also a typical 1D spin-gapped system. Fascinating phenomena have been investigated intensively using this model.  
It was shown numerically that there appears a plateau region on the magnetization curve at half of the saturation value \cite{tone1,tone2}. The phase diagram was determined precisely by the level spectroscopy analysis \cite{tone2}. 
Using bosonization technique and a perturbation calculation, a simple picture of the half-magnetization-plateau state is presented as the twofold degenerate state with the singlet and triplet pairs occupying the strong bonds alternately \cite{Totsuka}.

This model can be regarded as a minimal model for the spin-Peierls material $\rm{CuGeO_3}$ \cite{CCE,RD}. Note that the parameter sets proposed for $\rm{CuGeO_3}$ \cite{CCE,RD} lie within the half-magnetization-plateau region of the phase diagram \cite{tone2}. 
It was shown for these parameters that in $H_{c_1} <H$ the static spin susceptibility parallel to magnetic fields takes the maximum at the incommensurate (IC) wave vector \cite{PRH}. Using the adequate parameters for $\rm{CuGeO_3}$ \cite{CCE}, the critical exponents of the spin correlation functions were further investigated numerically in $H_{c_1} <H < H_{c_2}$ \cite{US1}. It was shown that the leading contribution of the spin correlation function changes depending on magnetic fields: In the middle range between $H_{c_1} <H < H_{c_2}$ the IC spin correlation parallel to magnetic fields becomes dominant, while around $H=H_{c_1}$ and $H=H_{c_2}$ the staggered spin correlation perpendicular to magnetic fields becomes dominant. Such critical properties may have a relation to the IC phase of ${\rm CuGeO_3}$ in magnetic fields \cite{sl}. 
Furthermore, the results suggest that the IC long-range order may be stabilized, when the interchain interaction is relevant.

In this paper, we investigate critical properties of a $S=1/2$ bond-alternating spin chain with a NNN interaction in magnetic fields. We turn our attention to the dominant longitudinal IC spin correlation. We then investigate the characteristics of the ordered state, when the interchain interaction is taken into account. 
In Sec. II, we calculate critical exponents of the spin correlation functions systematically in various parameter sets, combining a numerical diagonalization method with finite-size-scaling analysis based on conformal field theory \cite{ST1}. On the basis of the results, we determine the phase diagram for the dominant longitudinal IC spin correlation and discuss the origin of such properties. 
In Sec. III, we next investigate the ordered state in magnetic fields, by taking account of the interchain interaction. Combining the density-matrix renormalization group (DMRG) method with the interchain mean-field theory \cite{ST2,Schulz,Wang,Sand,WH,Sakai2,Kawa}, we calculate the staggered susceptibility and the uniform magnetization. We discuss whether the staggered long-range order can be stabilized in $H_{c_1} <H < H_{c_2}$. To investigate characteristics of the ordered state, we calculate the static structure factor.  
In Sec. IV, we discuss the recent experimental results for $1/T_1$ in magnetic fields performed for pentafluorophenyl nitronyl nitroxide (${\rm F_5PNN}$) \cite{Izumi}. 
Sec. V is devoted to the summary.

%%%%%%%%%%%%%%%%%%%%%%%%%%%%%%%%%%%%%%%%%%%%%%%%%%%%%%%%%%%%%%%%%%%%%%%%%
\section{Critical properties}\label{sec2}
%%%%%%%%%%%%%%%%%%%%%%%%%%%%%%%%%%%%%%%%%%%%%%%%%%%%%%%%%%%%%%%%%%%%%%%%%
\subsection{Model and method}
%%%%%%%%%%%%%%%%%%%%%%%%%%%%%%%%%%%%%%%%%%%%%%%%%%%%%%%%%%%%%%%
Let us first consider the 1D $S=1/2$ bond-alternating spin system with a NNN interaction in magnetic fields described by the following Hamiltonian,
%
%%%%%%%%%%%%%%%%%%%%%%%%%%%%%%%%%%%%%%%%%%%%%%%%%%%%%%%%%%%%%%%%%%%%
\begin{eqnarray}
{\mathcal H} = {\mathcal H_0} + {\mathcal H_{\rm Z}}, 
\label{eqn:ham1} 
\end{eqnarray}
\begin{eqnarray}
{\mathcal H_0}  &=& 2J \sum_{i=1}^{L} 
\bigg\{
     \left( 1+\delta \right)  
      \mbox{\boldmath$S$}_{i,l} \cdot \mbox{\boldmath$S$}_{i,r}   
  +  \left( 1-\delta \right)  
      \mbox{\boldmath$S$}_{i,r} \cdot \mbox{\boldmath$S$}_{i+1,l} 
  \nonumber \\
 &+&  \alpha \left[
      \mbox{\boldmath$S$}_{i,l} \cdot \mbox{\boldmath$S$}_{i+1,l} 
    + \mbox{\boldmath$S$}_{i,r} \cdot \mbox{\boldmath$S$}_{i+1,r} 
      \right]
\bigg\} ,
\label{eqn:ham2} 
\end{eqnarray}
\begin{eqnarray}
{\mathcal H_{\rm Z}} = - g\mu_B H \sum_{i=1}^{L} 
                      \left( S_{i,l}^{z} + S_{i,r}^{z} \right), 
\label{eqn:ham3} 
\end{eqnarray}
%%%%%%%%%%%%%%%%%%%%%%%%%%%%%%%%%%%%%%%%%%%%%%%%%%%%%%%%%%%%%%%%%%%%
%
where $L$ is the total number of unit cells, which consist of neighboring two spins, $\mbox{\boldmath$S$}_{i,l(r)}$ is the $S=1/2$ spin operator of the left-(right-) hand side in the $i$th unit cell, and $H$ is the magnitude of the external magnetic field along the $z$ axis. Here, $\delta$ is the bond-alternation parameter with $0<\delta<1$ and $2\alpha J$ is the NNN antiferromagnetic coupling. 
We set $J=g\mu_B=1$. The periodic boundary condition is applied. 

Since the system has translational symmetry and rotational symmetry about the $z$ axis, we can classify the Hamiltonian into the subspace according to the wave vector $k$ and the magnetization $M=\sum_{i=1}^{L} \left( S_{i,l}^{z} + S_{i,r}^{z} \right)$.  
The distance between the neighboring two unit cells is set to unity. Thus, the wave vector takes the discrete value $k=(2\pi/L)\times {\rm integer}$ for finite $L$.
Using Lanczos algorithm, the lowest energy in each subspace is calculated numerically. For the $2L$-spin system, we define the lowest energy of ${\mathcal H_0}$ in the magnetization $M$ and the wave vector $k$ as $E_{k}(L,M)$.  
In given $L$ and $M$, $E_{k}(L,M)$ takes the minimum at $k=k_0$. We simply describe $E_{k_0}(L,M)$ as $E(L,M)$. 
Following the method developed in Ref. 3, we investigate critical properties of the system in magnetic fields.

%%%%%%%%%%%%%%%%%%%%%%%%%%%%%%%%%%%%%%%%%%%%%%%%%%%%%%%%%%%%%%%
\subsection{Central charge}
%%%%%%%%%%%%%%%%%%%%%%%%%%%%%%%%%%%%%%%%%%%%%%%%%%%%%%%%%%%%%%%
We first investigate the central charge $c$ by use of 
%
%%%%%%%%%%%%%%%%%%%%%%%%%%%%%%%%%%%%%%%%%%%%%%%%%%%%%%%%%%%%%%%%%%%%
\begin{eqnarray}
\frac{1}{L}E(L,M) \sim \epsilon(m) - \frac{\pi cv_{\rm s}}{6}\frac{1}{L^2}, 
\label{eqn:ch} 
\end{eqnarray}
%%%%%%%%%%%%%%%%%%%%%%%%%%%%%%%%%%%%%%%%%%%%%%%%%%%%%%%%%%%%%%%%%%%%
where $\epsilon(m)$ is the ground state energy per a unit cell in the thermodynamic limit with $m=M/(2L)$ $(0\leq m \leq1/2)$.  
The velocity is estimated as 
%
%%%%%%%%%%%%%%%%%%%%%%%%%%%%%%%%%%%%%%%%%%%%%%%%%%%%%%%%%%%%%%%%%%%%
\begin{eqnarray}
v_{\rm s} = \frac{L}{2\pi} \left[E_{k_1}(L,M)-E(L,M)\right], 
\label{eqn:vel} 
\end{eqnarray}
%%%%%%%%%%%%%%%%%%%%%%%%%%%%%%%%%%%%%%%%%%%%%%%%%%%%%%%%%%%%%%%%%%%%
where $k_1$ is the wave vector closest to $k_0$. 
From the $L$ dependence of $E(L,M)/L$, we derive $(\pi/6)cv_{\rm s}$.  Combining $(\pi/6)cv_{\rm s}$ thus obtained and numerically calculated $v_{\rm s}$, we can evaluate the central charge $c$ numerically. 
Typical results are shown in Table I. 
From the results, we conclude that $c=1$ for $\alpha = 0.10, 0.15, 0.18$ and $0.20$ in $\delta=0.15$ and the system in the gapless region can be described as the Tomonaga-Luttinger (TL) liquid. 

%%%%%%%%%%%%%%%%%%%%%%%%%%%%%%%%%%%%%%%%%%%%%%%%%%%%%%%%%%%%%%%%%%%%%%%%%%%%
%                          TABLE I                                         %
%%%%%%%%%%%%%%%%%%%%%%%%%%%%%%%%%%%%%%%%%%%%%%%%%%%%%%%%%%%%%%%%%%%%%%%%%%%%
% 
\begin{table}[htb]
\caption{
The central charge for $\alpha = 0.10, 0.15, 0.18$ and $0.20$ in $\delta=0.15$. 
}
\begin{tabular}{rccccc} 
\hline \hline 
 &$\alpha$ & 0.10 & 0.15 & 0.18 & 0.20 \\ 
\hline
$m=$&$1/8$& 
1.10 & 1.10 & 1.10 & 1.12 \\ 
 &$1/7$&  
1.08 & 1.06 & 1.09 & 1.09 \\ 
 &$1/6$&  
1.07 & 1.06 & 1.08 & 1.08 \\
 &$1/5$&  
1.09 & 1.09 & 1.09 & 1.09 \\  
 &$3/14$ & 
1.06 & 1.06 & 1.06 & 1.06 \\ 
 &$1/4$ & 
1.06 & 1.06 & 1.05 & 1.05 \\ 
 &$2/7$ & 
1.05 & 1.06 & 1.05 & 1.05 \\ 
 &$3/10$ & 
1.07 & 1.06 & 1.05 & 1.04 \\ 
 &$1/3$ & 
1.06 & 1.06 & 1.05 & 1.05 \\ 
 &$5/14$ & 
1.06 & 1.00 & 1.03 & 1.02 \\ 
 &$3/8$ & 
1.07 & 1.06 & 1.05 & 1.03 \\ 
 &$2/5$ & 
1.07 & 1.06 & 1.05 & 1.03 \\ 
\hline \hline 
\end{tabular}
\end{table}
%%%%%%%%%%%%%%%%%%%%%%%%%%%%%%%%%%%%%%%%%%%%%%%%%%%%%%%%%%%%%%%%%%%%%%%%%%%%

It was shown that in the adequate parameters of the Hamiltonian (\ref{eqn:ham1}) the plateau region appears at half of the saturation value ($m=1/4$) in the magnetization curve  \cite{tone1,tone2,Totsuka}. 
We investigate the magnetization curve using the asymptotic forms of the excitation energy 
%
%%%%%%%%%%%%%%%%%%%%%%%%%%%%%%%%%%%%%%%%%%%%%%%%%%%%%%%%%%%%%%%%%%%%
\begin{eqnarray}
E(L,M+1)-E(L,M)-H^+(m) \sim \pi v_{\rm s}\eta^x \frac{1}{L}, 
\label{eqn:as1} 
\end{eqnarray}
\begin{eqnarray}
E(L,M)-E(L,M-1)-H^-(m) \sim -\pi v_{\rm s}\eta^x \frac{1}{L}, 
\label{eqn:as2} 
\end{eqnarray}
%%%%%%%%%%%%%%%%%%%%%%%%%%%%%%%%%%%%%%%%%%%%%%%%%%%%%%%%%%%%%%%%%%%%
where $\eta^x$ is the critical exponent of the spin correlation function transverse to the magnetic field, $H^+(m)$ and $H^-(m)$ are the magnetic field in a given $m$ in the thermodynamic limit. 
If $H^+(m_0)=H^-(m_0)$, no plateau appears at $m=m_0$ in the magnetization curve.

In the parameters used in Table I, $H^+(m)$ and $H^-(m)$ extrapolated into $L \rightarrow \infty$ satisfy $H^+(m)=H^-(m)$ within the accuracy of $O(10^{-4})$. Therefore, we conclude that no plateau appears at $m=1/4$ within our numerical accuracy.

%%%%%%%%%%%%%%%%%%%%%%%%%%%%%%%%%%%%%%%%%%%%%%%%%%%%%%%%%%%%%%%
\subsection{Critical exponents}
%%%%%%%%%%%%%%%%%%%%%%%%%%%%%%%%%%%%%%%%%%%%%%%%%%%%%%%%%%%%%%%
We next investigate the critical exponents of the spin correlation functions. 
The spin correlation functions of the long-distance behavior in the TL liquid take the forms as 
%
%%%%%%%%%%%%%%%%%%%%%%%%%%%%%%%%%%%%%%%%%%%%%%%%%%%%%%%%%%%%%%%%%%%%
\begin{eqnarray}
\langle \phi^z_n \phi^z_1 \rangle \sim m^2 + C_1 n^{-2} 
                                     + C_2 n^{-\eta^z} \cos[2k_{\rm F}n] , 
\label{eq:rdsl}
\end{eqnarray}
\begin{eqnarray}
\langle \phi^x_n \phi^x_1 \rangle \sim D_1 n^{-\eta^x} \cos[\pi n], 
\label{eqn:rdst}
\end{eqnarray}
%%%%%%%%%%%%%%%%%%%%%%%%%%%%%%%%%%%%%%%%%%%%%%%%%%%%%%%%%%%%%%%%%%%%
%
where $\phi^{z(x)}_n$ is composed of the two spin operators at the $n$th unit cell and $2k_{\rm F}=\pi(1-2m)$. 
The critical exponents are given by 
%
%%%%%%%%%%%%%%%%%%%%%%%%%%%%%%%%%%%%%%%%%%%%%%%%%%%%%%%%%%%%%%%%%%%%
\begin{equation}
\eta^{z}=\frac{L}{\pi v_{\rm s}} \left[E_{2k_{\rm F}}(L,M)-E(L,M)\right], 
\label{eqn:exp1}
\end{equation}
\begin{equation}
\eta^{x}=\frac{L}{2\pi v_{\rm s}} \left[E(L,M+1)+E(L,M-1)-2E(L,M)\right]. 
\label{eqn:exp2}
\end{equation}
%%%%%%%%%%%%%%%%%%%%%%%%%%%%%%%%%%%%%%%%%%%%%%%%%%%%%%%%%%%%%%%%%%%%
%
Using the expressions (\ref{eqn:exp1}) and (\ref{eqn:exp2}), we calculate $\eta^x$ and $\eta^z$. Since the size dependence of $\eta^x$ and $\eta^z$ is found to be well fitted by $O(L^{-2})$, we extrapolate the results in a given $m$ to $L \rightarrow \infty$. 

%%%%%%%%%%%%%%%%%%%%%%%%%%%%%%%%%%%%%%%%%%%%%%%%%%%%%%%%%%%%%%%%%%%%%%%%%%%%
%                          TABLE II                                        %
%%%%%%%%%%%%%%%%%%%%%%%%%%%%%%%%%%%%%%%%%%%%%%%%%%%%%%%%%%%%%%%%%%%%%%%%%%%%
% 
\begin{table}[htb]
\caption{
The value $\eta^x \cdot \eta^z$ for $\alpha = 0.10, 0.15, 0.18$ and $0.20$ in $\delta=0.15$. 
}
\begin{tabular}{rccccc} 
\hline \hline 
 &$\alpha$ & 0.10 & 0.15 & 0.18 & 0.20 \\ 
\hline
$m=$&$1/8$& 
0.98 & 0.99 & 0.98 & 0.97 \\ 
 &$1/7$ & 
0.99 & 0.99 & 0.99 & 0.99 \\ 
 &$1/6$ & 
1.00 & 1.00 & 1.00 & 1.00 \\
 &$1/5$ & 
1.00 & 1.00 & 0.99 & 0.99 \\  
 &$3/14$ & 
1.00 & 1.00 & 1.00 & 1.00 \\ 
 &$1/4$ & 
0.99 & 0.96 & 0.93 & 0.91 \\ 
 &$2/7$ & 
1.00 & 1.00 & 1.00 & 0.99 \\ 
 &$3/10$ & 
1.00 & 0.99 & 0.99 & 0.99 \\ 
 &$1/3$ & 
1.00 & 1.00 & 1.00 & 1.00 \\ 
 &$5/14$ & 
0.99 & 1.00 & 1.00 & 0.99 \\ 
 &$3/8$ & 
1.00 & 1.00 & 0.99 & 0.99 \\ 
 &$2/5$ & 
1.00 & 1.00 & 1.00 & 0.99 \\ 
\hline \hline 
\end{tabular}
\end{table}
%%%%%%%%%%%%%%%%%%%%%%%%%%%%%%%%%%%%%%%%%%%%%%%%%%%%%%%%%%%%%%%%%%%%%%%%%%%%
%%%%%%%%%%%%%%%%%%%% Fig. 1 %%%%%%%%%%%%%%%%%%%%%%%%%%%%%%%%%%
\begin{figure}[h]
\begin{center}
\includegraphics[trim=20mm 10mm 10mm 30mm,clip,scale=0.45]{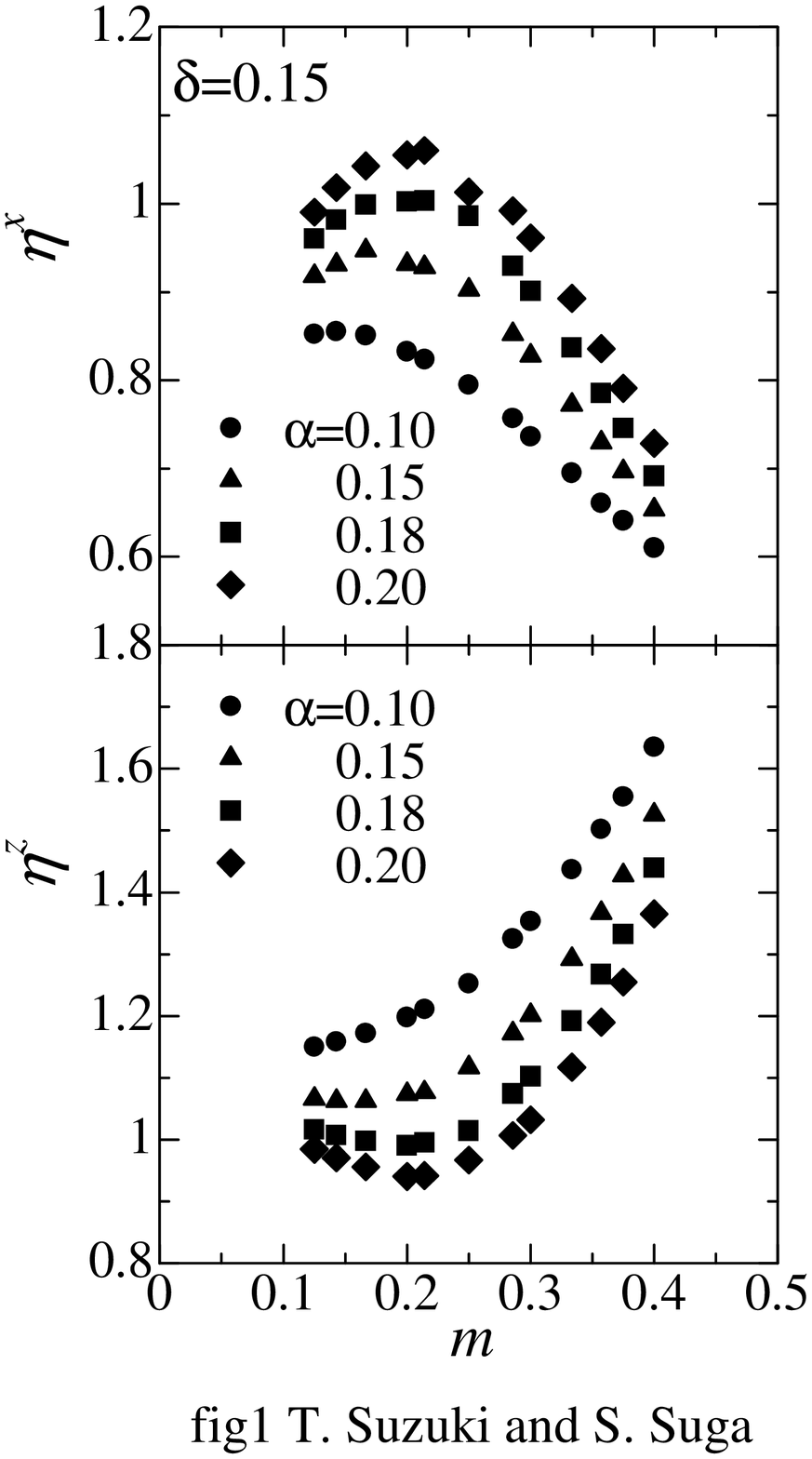}
\end{center}
\vspace{-7mm}
\caption{\small
The extrapolated $\eta^x$ and $\eta^z$ for $\alpha = 0.10, 0.15, 0.18$ and $0.20$ in $\delta=0.15$ as function of $m$.
}
\label{fig:fig1}
\end{figure}
%%%%%%%%%%%%%%%%%%%%%%%%%%%%%%%%%%%%%%%%%%%%%%%%%%%%%%%%%%%%%

In Fig. 1, we show the extrapolated $\eta^x$ and $\eta^z$ for $\alpha = 0.10, 0.15, 0.18$ and $0.20$ in $\delta=0.15$ as function of $m$. 
It is a difficult issue to obtain the precise results in $m<0.1$, because we cannot treat larger size systems. At the lower critical field and the saturation field, the system may be described by a boson with infinitely large repulsion. Therefore, the critical exponents take the values $\eta^z=2$ and $\eta^x=1/2$ at $m=0$ and $1/2$.  
On the basis of these results, we conclude that in $\alpha=0.10$, and $0.15$, the relation $\eta^x <1< \eta^z$ is satisfied in $0 \leq m \leq 1/2$. The results indicate that the transverse staggered spin correlation is dominant in $0 \leq m \leq 1/2$. 
In $\alpha=0.18$ and $0.20$, on the other hand, there appears the region where the relation $\eta^z < 1 < \eta^x$ is satisfied in $0.16 <m< 0.22$ and $0.13 <m< 0.28$, respectively. 
The results indicate that the dominant spin correlation changes depending on the magnetization: For $\alpha=0.18$ the longitudinal IC spin correlation becomes dominant in $0.16 <m< 0.22$, while the transverse staggered spin correlation becomes dominant in $0 \leq m <0.16$ and $0.23 < m \leq 1/2$. 
For $\alpha=0.20$ the longitudinal IC spin correlation becomes dominant in $0.13 <m< 0.28$, while the transverse staggered spin correlation becomes dominant in $0 \leq m <0.13$ and $0.28 < m \leq 1/2$. 
To check the numerical accuracy, we evaluate the value $\eta^x \cdot \eta^z$. From the results shown in Table II, the universal relation $\eta^x \cdot \eta^z = 1$ characteristic of the TL liquid \cite{LP,Hal} is well satisfied in $0<m<1/2$ except for $m=1/4$ at $\alpha=0.15, 0.18$, and $0.20$.

The appearance of the half-magnetization plateau in this system is the Berezinskii-Kosterlitz-Thouless quantum phase transition \cite{tone2}. Therefore, in finite systems, $\eta^x$ and $\eta^z$ at $m=1/4$ are suffering from slowly-converging logarithmic size-corrections in the vicinity of the transition point. As will be seen in Fig. 2, the parameters $\alpha=0.15, 0.18$, and $0.20$ in $\delta=0.15$ lie close to the transition point. It is thus difficult to obtain their highly accurate results for $\eta^x$ and $\eta^z$ at $m=1/4$.

We develop the calculation in other parameter sets of $\alpha$ and $\delta$, turning our attention to the behavior $\eta^z <1< \eta^x$. The results are summarized in Fig. 2. 
In the darkly shaded area, the relation $\eta^z <1< \eta^x$ is satisfied around the half-magnetization of the saturation. In the lightly shaded area, the relation $\eta^x <1< \eta^z$ is satisfied in $0<m<1/2$. The half-magnetization plateau appears in the `plateau' area. 
The two broken lines are the boundaries for the half-magnetization plateau obtained by the level spectroscopy analysis \cite{tone2}.  Note that the method used in this paper is difficult to detect the small half-magnetization plateau close to the broken lines \cite{comm}. 

%%%%%%%%%%%%%%%%%%%% Fig. 2 %%%%%%%%%%%%%%%%%%%%%%%%%%%%%%%%%%
\begin{figure}[h]
\begin{center}
\includegraphics[trim=12mm 80mm 5mm 25mm,clip,scale=0.45]{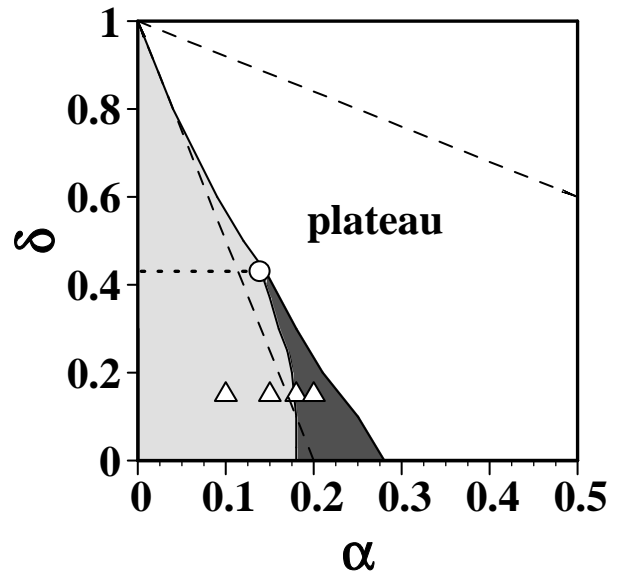}
\end{center}
\vspace{-7mm}
\caption{\small
In the darkly shaded area, the relation $\eta^z <1< \eta^x$ is satisfied around the half-magnetization of the saturation. In the lightly shaded area, the relation $\eta^x <1< \eta^z$ is satisfied in $0<m<1/2$. The half-magnetization plateau appears in the `plateau' area. Two broken lines are the boundaries for the half-magnetization plateau obtained by the level spectroscopy analysis \cite{tone2}. The open circle corresponds to the parameters for ${\rm F_5PNN}$ ($\alpha=0.15$ and $\delta=0.43$). The four open triangles correspond to the parameters used in Fig. 1 ($\alpha = 0.10, 0.15, 0.18$ and $0.20$ in $\delta=0.15$). 
}
\label{fig:fig2}
\end{figure}
%%%%%%%%%%%%%%%%%%%%%%%%%%%%%%%%%%%%%%%%%%%%%%%%%%%%%%%%%%%%%

Such a dominant longitudinal IC spin correlation is closely related to the formation of the half-magnetization plateau, which can be regarded as the $4k_{\rm F}$ CDW state \cite{Totsuka}.  
It was shown that in the vicinity of the CDW transition point the dressed charge is suppressed and takes the value typical of the strongly correlated system \cite{Hub1,Hub2,Hub3,MZ,Poil}. 
In the 1D Hubbard model, for example, the dressed charge of the charge excitation takes the value for the spinless fermion at half-filling irrespective of the strength of the Coulomb repulsion \cite{Hub1,Hub2,Hub3}. 
Within the method used in this paper, it is difficult to investigate the critical properties close to the half-magnetization plateau. 
To see the critical properties in the other viewpoint, we calculate $\eta^z$ in $\alpha \sim 0$ and $\delta \sim 1$ on the basis of the effective model.

%%%%%%%%%%%%%%%%%%%%%%%%%%%%%%%%%%%%%%%%%%%%%%%%%%%%%%%%%%%%%%%
\subsection{Critical properties based on the effective model}
%%%%%%%%%%%%%%%%%%%%%%%%%%%%%%%%%%%%%%%%%%%%%%%%%%%%%%%%%%%%%%%
Under the condition $\alpha \sim 0$ and $\delta \sim 1$, the Hamiltonian (\ref{eqn:ham1}) can be mapped onto the 1D $S=1/2 \, XXZ$ model in effective magnetic fields \cite{Totsuka,mila,GT,FZ}:  
% 
%%%%%%%%%%%%%%%%%%%%%%%%%%%%%%%%%%%%%%%%%%%%%%%%%%%%%%%%%%%%%%%%%%%%
\begin{eqnarray}
{\mathcal H}_{\rm eff}
 &=& J_{\rm eff} \sum_{i} \left[ 
       \tilde{S}_{i}^{x} \tilde{S}_{i+1}^{x}  
                + \tilde{S}_{i}^{y} \tilde{S}_{i+1}^{y} 
     + \Delta \tilde{S}_{i}^{z} \tilde{S}_{i+1}^{z} 
      \right]    \nonumber \\
 &-& H_{\rm eff} \sum_{i} \tilde{S}_{i}^{z}, 
\label{eqn:hameff}
\end{eqnarray}
%%%%%%%%%%%%%%%%%%%%%%%%%%%%%%%%%%%%%%%%%%%%%%%%%%%%%%%%%%%%%%%%%%%%
%
where $\tilde{S}^{a}_{i} \, (a=x, y, z)$ is the pseudo spin operator made up of the two states of the two spins on the strong bond. 
The effective coupling constants and the effective magnetic field are given by using the original parameters as 
$J_{\rm eff}=[2\alpha - (1-\delta)]$, 
$\Delta J_{\rm eff} =[\alpha + \frac{1}{2}(1-\delta)]$ and 
$H_{\rm eff}=H-[\alpha+\frac{1}{2}(5+3\delta)]$.

When $\Delta >1$ with $J_{\rm eff}>0$, the ground state has N\'{e}el order and the excitations are gapped. Since the magnetization of the effective Hamiltonian $m_{\rm eff}$ satisfies the relation $m_{\rm eff}+1/2=2m$, the ground state $m_{\rm eff}=0$ corresponds to the half-magnetization-plateau state of the Hamiltonian (\ref{eqn:ham1}). 
The critical exponents of the spin correlation functions in ${\mathcal H}_{\rm eff}$ are the same as those in the Hamiltonian (\ref{eqn:ham1}) \cite{GT,FZ,HS}. Therefore, we calculate the critical exponents in ${\mathcal H}_{\rm eff}$ to see the critical properties around $\alpha \sim 0$ and $\delta \sim 1$ of the Hamiltonian (\ref{eqn:ham1}). 

%%%%%%%%%%%%%%%%%%%% Fig. 3 %%%%%%%%%%%%%%%%%%%%%%%%%%%%%%%%%%
\begin{figure}[h]
\begin{center}
\includegraphics[trim=20mm 45mm 10mm 5mm,clip,scale=0.45]{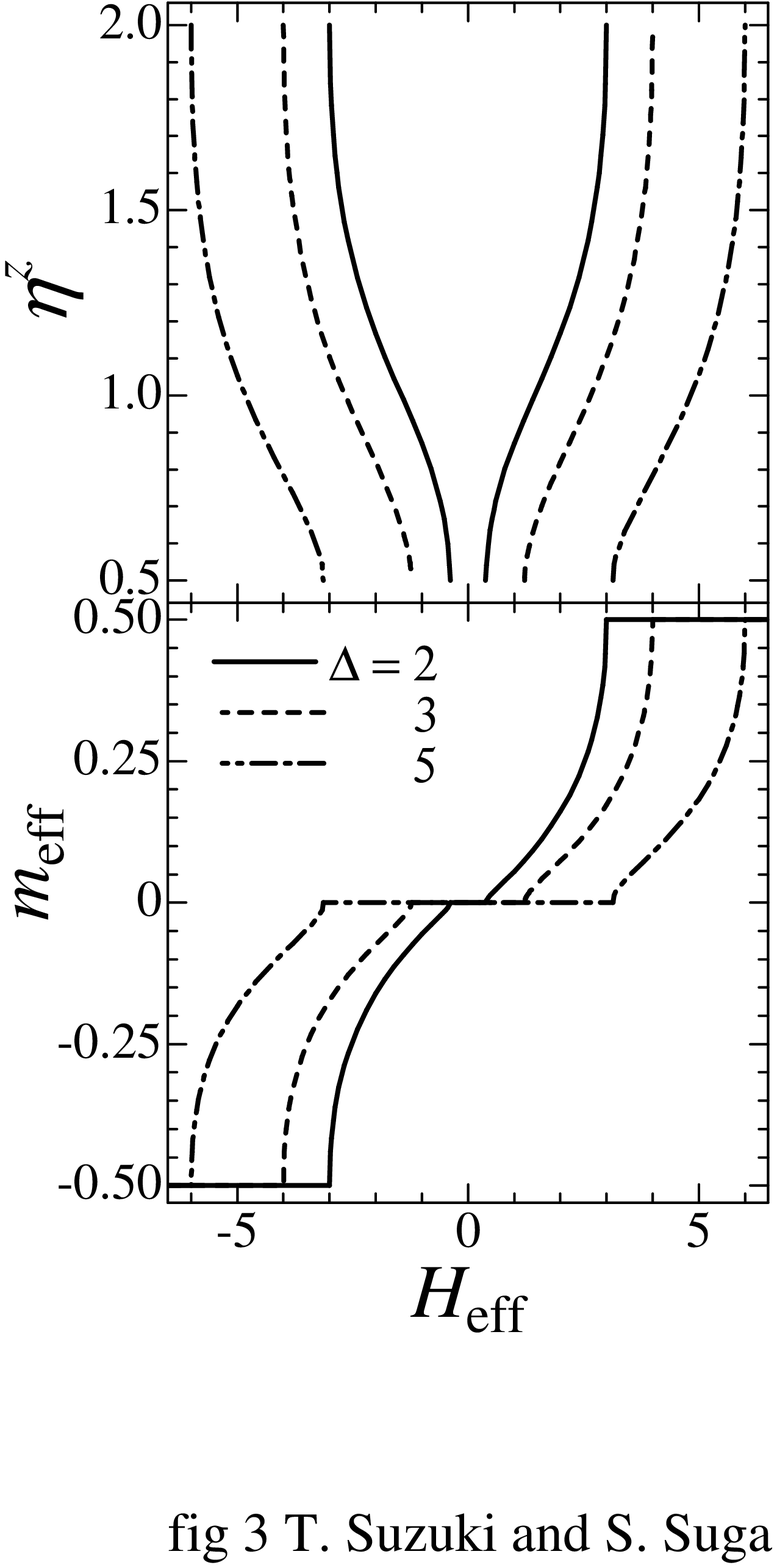}
\end{center}
\vspace{-7mm}
\caption{\small
$H_{\rm eff}$ dependence of $\eta^z$ and $m_{\rm eff}$ for several $\Delta$. 
}
\label{fig:fig3}
\end{figure}
%%%%%%%%%%%%%%%%%%%%%%%%%%%%%%%%%%%%%%%%%%%%%%%%%%%%%%%%%%%%%%

The effective Hamiltonian enables us to calculate $\eta^z$ exactly in the gapless region using the Bethe ansatz solution \cite{text}. 
The critical exponent $\eta^z$ is obtained from the dressed charge as $\eta^z = 2[Z(\Lambda)]^2$ \cite{BIK}. The dressed charge $Z(\lambda)$ is obtained from the integral equation, 
%
%%%%%%%%%%%%%%%%%%%%%%%%%%%%%%%%%%%%%%%%%%%%%%%%%%%%%%%%%%%%%%%%%%%%
\begin{eqnarray}
Z(\lambda) = 1 - \frac{1}{2\pi} \int_{-\Lambda}^{\Lambda} \, d\lambda^{\prime} 
             a_2(\lambda-\lambda^{\prime}) Z(\lambda^{\prime}), 
\label{eqn:dc}
\end{eqnarray}
%%%%%%%%%%%%%%%%%%%%%%%%%%%%%%%%%%%%%%%%%%%%%%%%%%%%%%%%%%%%%%%%%%%%
%
where $a_n(\lambda) = \phi \sinh(n\phi)/[\cosh(n\phi) - \cos(\phi \lambda)]$ 
with $\Delta=\cosh \phi$ $(\phi>0)$. 
The cutoff $\Lambda$ is determined by the condition for the dressed energy $\varepsilon(\pm\Lambda)=0$, where $\varepsilon(\lambda)$ is obtained from the integral equation in given magnetic field as 
%
%%%%%%%%%%%%%%%%%%%%%%%%%%%%%%%%%%%%%%%%%%%%%%%%%%%%%%%%%%%%%%%%%%%%
\begin{eqnarray}
\varepsilon(\lambda) = \varepsilon_0(\lambda) - 
  \frac{1}{2\pi} \int_{-\Lambda}^{\Lambda} \, d\lambda^{\prime} 
             a_2(\lambda-\lambda^{\prime}) \varepsilon(\lambda^{\prime}) 
\label{eqn:de}
\end{eqnarray}
%%%%%%%%%%%%%%%%%%%%%%%%%%%%%%%%%%%%%%%%%%%%%%%%%%%%%%%%%%%%%%%%%%%%
%
with 
$\varepsilon_0(\lambda) = 
                 H_{\rm eff} - [2\pi \tilde{J}\sinh \phi/\phi] a_1(\lambda)$. 
The magnetization is obtained by use of the dressed charge as 
%
%%%%%%%%%%%%%%%%%%%%%%%%%%%%%%%%%%%%%%%%%%%%%%%%%%%%%%%%%%%%%%%%%%%%
\begin{eqnarray}
m_{\rm eff} = \frac{1}{2} - \frac{1}{2\pi} \int_{-\Lambda}^{\Lambda} \, d\lambda 
a_1(\lambda) Z(\lambda).
\label{eqn:mg}
\end{eqnarray}
%%%%%%%%%%%%%%%%%%%%%%%%%%%%%%%%%%%%%%%%%%%%%%%%%%%%%%%%%%%%%%%%%%%%
%
We calculate $\eta^z$ and $m_{\rm eff}$ as function of $H_{\rm eff}$. The results are shown in Fig. 3 for several anisotropic parameters $\Delta$. 
Around both ends of the half-magnetization plateau, the relation $\eta^z < 1$ is satisfied and the longitudinal IC spin correlation becomes dominant. 
It is calculated analytically that $\eta^z=1/2$ at the lower critical fields $H_{\rm eff,c_1}=\pm [2\sinh(\phi)/\phi] K(m^{\prime})m^{\prime}$ and $\eta^z=2$ at the saturation fields $H_{\rm eff,c_2}=\pm (1+\Delta)$ irrespective of $\Delta (>1)$, where $K(m^{\prime})$ is the complete elliptic integral and its modulus $m^{\prime}$. 
The half-magnetization-plateau state in the Hamiltonian (\ref{eqn:ham1}) is N\'{e}el ordered in the language of the pseudo spin. 
As mentioned before, the dressed charge is suppressed in the vicinity of the CDW transition point. The results that $\eta^z$ takes the minimum at $H_{\rm eff}=H_{\rm eff,c_1}$ can be explained in this point of view.

Judging from these findings, we conclude that such a dominant longitudinal IC spin correlation emerges around the half-magnetization-plateau region.

%%%%%%%%%%%%%%%%%%%%%%%%%%%%%%%%%%%%%%%%%%%%%%%%%%%%%%%%%%%%%%%
\subsection{NMR relaxation rate}
%%%%%%%%%%%%%%%%%%%%%%%%%%%%%%%%%%%%%%%%%%%%%%%%%%%%%%%%%%%%%%%
We have shown that the longitudinal IC spin correlation becomes dominant around the half-magnetization-plateau region in the system described by the Hamiltonian (\ref{eqn:ham1}). Experimentally, such a feature can be observed by the NMR relaxation rate $1/T_1$. 
When the NMR is done on the nuclei located at the different sites from the electronic spins, the relaxation occurs through a dipolar interaction between the nuclear and electronic spins. 
In this case, $1/T_1$ of the TL liquid is expressed as a sum of contributions from the longitudinal and transverse dynamical spin susceptibilities. 
Paying attention to the divergence behavior of $1/T_1$, we obtain the expression \cite{HS} $1/T_1 \sim \tilde{C}_2 T^{\eta^z-1} + \tilde{D}_1 T^{\eta^x-1}$, 
where the first term originates in the longitudinal IC spin susceptibility and the second one originates in the transverse staggered spin susceptibility.
Note that $\tilde{C}_2$ and $\tilde{D}_1$ depend on temperature and magnetic field. However, their effects are probably weaker than the divergence behavior. 
Therefore, we consider only the field dependence of the divergence exponent of $1/T_1$. Depending on $\eta^x$ and $\eta^z$, $1/T_1$ takes the form as
%
%%%%%%%%%%%%%%%%%%%%%%%%%%%%%%%%%%%%%%%%%%%%%%%%%%%%%%%%%%%%%%%%%%%%
\begin{equation}
\frac{1}{T_1} \sim \left\{
\begin{array}{lll}
T^{\eta^x -1} & \equiv T^{-\gamma}, & \eta^x<1<\eta^z \\
T^{\eta^z -1} & \equiv T^{-\gamma}, & \eta^z<1<\eta^x. 
\end{array}
\right.
\label{eqn:nmr1}
\end{equation}
%%%%%%%%%%%%%%%%%%%%%%%%%%%%%%%%%%%%%%%%%%%%%%%%%%%%%%%%%%%%%%%%%%%%
%
We discuss the divergence property of $1/T_1$ using the results for $\eta^x$ and $\eta^z$ shown in Fig. \ref{fig:fig1}. 
For $\alpha = 0.10$ and $0.15$ in $\delta=0.15$, the relation $\eta^x<1<\eta^z$ is satisfied in $0 < m < 1/2$, indicating that the divergence behavior is caused by the transverse staggered spin correlation. 
For $\alpha = 0.18$ and $0.20$ in $\delta=0.15$, the relation $\eta^z<1<\eta^x$ is satisfied in $0.16 <m< 0.22$ and $0.13 <m< 0.28$, respectively. In the following, we express these regions as $m_{p_1}<m<m_{p_2}$. Thus, for $\alpha = 0.18$ and $0.20$ in $\delta=0.15$, the relation $\eta^x<1<\eta^z$ is satisfied in $0 \leq m < m_{p_1}$ and $m_{p_2} < m \leq 1/2$. 
Therefore, in these parameters the divergence behavior is caused by the longitudinal IC spin correlation in $m_{p_1}<m<m_{p_2}$, and by the transverse staggered spin correlation in $0 \leq m<m_{p_1}$ and $m_{p_2}<m \leq 1/2$. 

%%%%%%%%%%%%%%%%%%%% Fig. 4 %%%%%%%%%%%%%%%%%%%%%%%%%%%%%%%%%%
\begin{figure}[h]
\begin{center}
\includegraphics[trim=25mm 110mm 20mm 45mm,clip,scale=0.45]{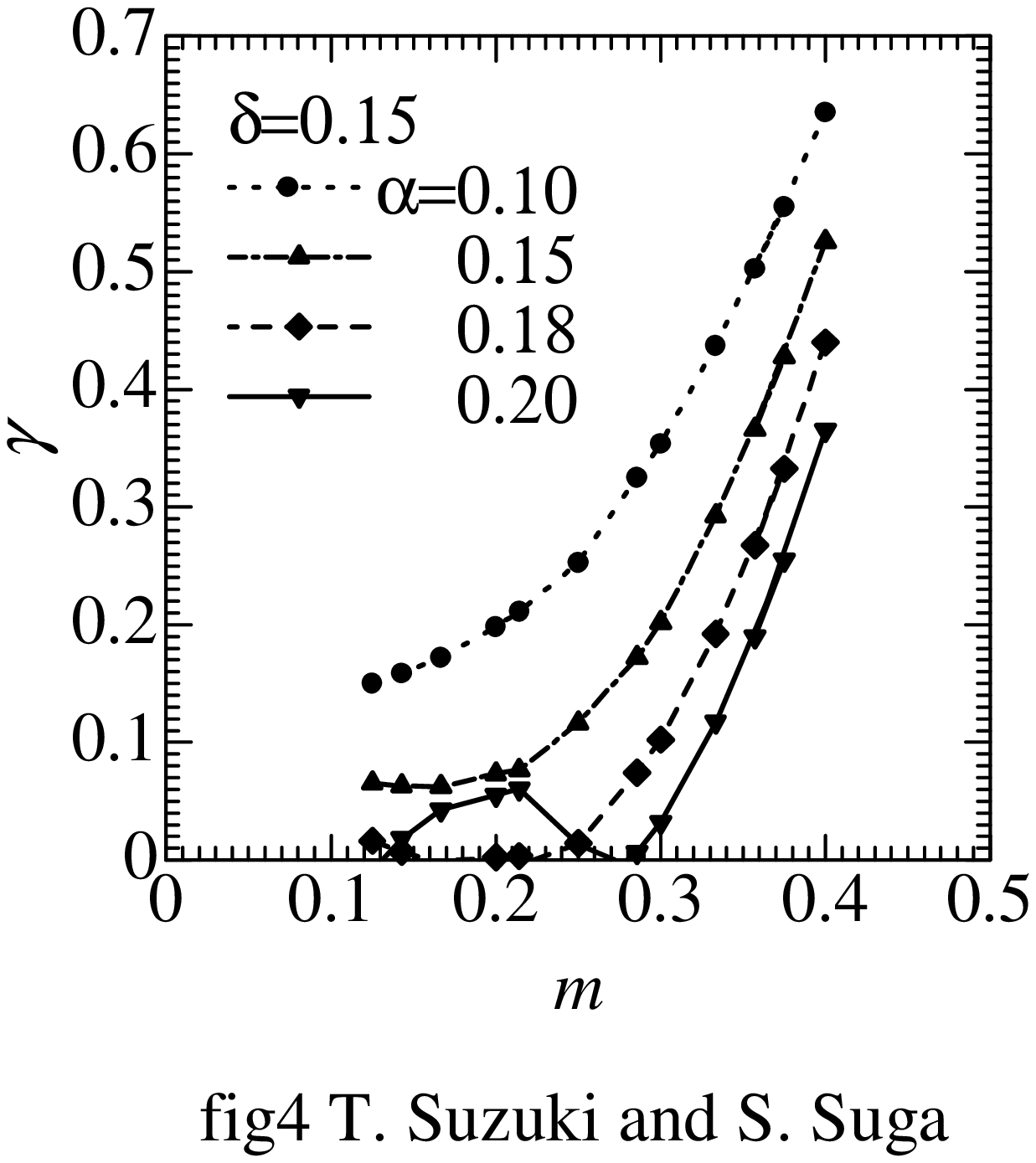}
\end{center}
\vspace{-7mm}
\caption{\small
The divergence exponents $\gamma$ for $\alpha = 0.10, 0.15, 0.18$ and $0.20$ in $\delta=0.15$ as function of $m$, where $1/T_1 \sim T^{-\gamma}$. 
The lines are to guide the eyes. 
}
\label{fig:fig4}
\end{figure}
%%%%%%%%%%%%%%%%%%%%%%%%%%%%%%%%%%%%%%%%%%%%%%%%%%%%%%%%%%%%%%

The $m$ dependence of the divergent exponent $\gamma$ is summarized in Fig. 4. 
For $\alpha = 0.10$ and $0.15$ in $\delta=0.15$, $\gamma$ varies concavely as function of $m$ with $\gamma=1/2$ at $m=0$ and $1/2$. 
For $\alpha = 0.18$ and $0.20$ in $\delta=0.15$, as $m$ increases, $\gamma$ decreases in $0 \leq m<m_{p_1}$, $\gamma$ varies convexly in $m_{p_1}<m<m_{p_2}$, and $\gamma$ increases in $m_{p_2}<m \leq 1/2$. It is noteworthy that $\gamma=0$ at $m=m_{p_1}$ and $m_{p_2}$, where $1/T_1$ shows no divergence and becomes almost independent of temperature. 
Such a feature can be observed experimentally. In fact, it was reported for ${\rm F_5PNN}$ that $1/T_1$ exhibits power-law behavior in $H_{c_1}<H<H_{c_2}$, and at a certain magnetic field $1/T_1$ became almost independent of temperature \cite{Izumi}. The results may be an evidence for the change of the dominant spin correlation in magnetic fields. Details will be discussed in Sec. IV.

%
%%%%%%%%%%%%%%%%%%%%%%%%%%%%%%%%%%%%%%%%%%%%%%%%%%%%%%%%%%%%%%%%%%%%%%%%%
\section{EFFECTS OF THE INTERCHAIN COUPLING}\label{sec4}
%%%%%%%%%%%%%%%%%%%%%%%%%%%%%%%%%%%%%%%%%%%%%%%%%%%%%%%%%%%%%%%%%%%%%%%%%
\subsection{Mean-field approximation}
%%%%%%%%%%%%%%%%%%%%%%%%%%%%%%%%%%%%%%%%%%%%%%%%%%%%%%%%%%%%%%%
We next consider the effects of an interchain interaction on the system described by the Hamiltonian (\ref{eqn:ham1}). The Hamiltonian may be written down as 
%
%%%%%%%%%%%%%%%%%%%%%%%%%%%%%%%%%%%%%%%%%%%%%%%%%%%%%%%%%%%%%%%%%%%%
\begin{eqnarray}
{\mathcal H} &=& 2J \sum_{i,j} \left\{ \left[1+(-1)^{i} \delta \right]  
  \mbox{\boldmath$S$}_{i,j} \cdot \mbox{\boldmath$S$}_{i+1,j}  
  + \alpha \mbox{\boldmath$S$}_{i,j} \cdot \mbox{\boldmath$S$}_{i+2,j} \right\}
   \nonumber \\
 &+& J^{\prime} \sum_{i,\langle j,j^{\prime}\rangle} 
  \mbox{\boldmath$S$}_{i,j} \cdot \mbox{\boldmath$S$}_{i,j^{\prime}} 
 - g\mu_B H \sum_{i,j} S_{i,j}^{z}, 
\label{eqn:ham4} 
\end{eqnarray}
%%%%%%%%%%%%%%%%%%%%%%%%%%%%%%%%%%%%%%%%%%%%%%%%%%%%%%%%%%%%%%%%%%%%
%
where $\mbox{\boldmath$S$}_{i,j}$ is the $S=1/2$ spin operator at the $i$th site in the $j$th chain, $J^{\prime}$ is the interchain antiferromagnetic coupling, and $\langle j,j^{\prime}\rangle$ denotes the summation over the pairs of nearest-neighbor chains. 
To put through the interchain mean-field treatment \cite{ST2,Schulz,Wang,Sand,WH,Sakai2,Kawa}, we introduce two kinds of mean fields induced by the interchain interaction and the external magnetic field as $\langle S^x_i \rangle=-(-1)^{i}m_{\rm s}$ and $\langle S^z_i \rangle=m_{\rm u}$. The former is the staggered magnetization and the latter is the uniform magnetization. The 1D mean-field Hamiltonian thus obtained are described as 
%
%%%%%%%%%%%%%%%%%%%%%%%%%%%%%%%%%%%%%%%%%%%%%%%%%%%%%%%%%%%%%%%%%%%%
\begin{eqnarray}
{\mathcal H_{\rm MF}} &=& 2J \sum_{i} \left\{ \left[1+(-1)^{i} \delta \right]  
  \mbox{\boldmath$S$}_{i} \cdot \mbox{\boldmath$S$}_{i+1}  
  + \alpha \mbox{\boldmath$S$}_{i} \cdot \mbox{\boldmath$S$}_{i+2} \right\}
   \nonumber \\
 &-& \left(g\mu_B H - h_{\rm u} \right) \sum_{i} S_{i}^{z} - h_{\rm s} \sum_{i} S_{i}^{x}, 
\label{eqn:ham5} 
\end{eqnarray}
%%%%%%%%%%%%%%%%%%%%%%%%%%%%%%%%%%%%%%%%%%%%%%%%%%%%%%%%%%%%%%%%%%%%
%
where $h_{\rm u}$ and $h_{\rm s}$ are the effective internal fields given by $h_{\rm u}=zJ^{\prime}m_{\rm u}$ and $h_{\rm s}=zJ^{\prime}m_{\rm s}$, respectively, with $z$ being the number of the adjacent chains. 
The effective field is also defined as $H_{\rm u}=g\mu_B H - h_{\rm u}$. 
We set $J=g\mu_B=1$.

%%%%%%%%%%%%%%%%%%%%%%%%%%%%%%%%%%%%%%%%%%%%%%%%%%%%%%%%%%%%%%%
\subsection{Staggered susceptibility}
%%%%%%%%%%%%%%%%%%%%%%%%%%%%%%%%%%%%%%%%%%%%%%%%%%%%%%%%%%%%%%%
On the basis of ${\mathcal H_{\rm MF}}$, we calculate the staggered magnetizations per a site $m_{\rm s}$ and the uniform magnetizations per a site $m_{\rm u}$ by means of the infinite-system DMRG method. 
It is known that the original infinite-system DMRG method devised by White \cite{White} may suffer from a problem that the system in magnetic fields is trapped at metastable states during the renormalization process. To avoid this problem, we use the modified infinite-system DMRG algorithm by adopting the recursion relation to the wave function \cite{mod1,mod2}.

In given $H_{\rm u}$, $m_{\rm s}$ can be obtained as function of $h_{\rm s}$, while in given $zJ^{\prime}$, $m_{\rm u}$ can be obtained as function of $H_{\rm u}$. 
The results are shown in Figs. 5 and 6. 
Note that for the pure 1D system $(J^{\prime}=0)$, in $\alpha=0.20$ and $\delta=0.15$ the longitudinal IC spin correlation becomes dominant $(\eta^z <1< \eta^x)$ around the half-magnetization of the saturation, while in $\alpha=0.05$ and $\delta=0.40$ the transverse staggered spin correlation becomes dominant $(\eta^x <1< \eta^z)$ in $0 \leq m \leq 1/2$. Solving the numerical results for $m_{\rm s}$ self-consistently with $h_{\rm s}=zJ^{\prime}m_{\rm s}$, we investigate whether the long-range staggered order exists or not. 

%%%%%%%%%%%%%%%%%%%% Fig. 5 %%%%%%%%%%%%%%%%%%%%%%%%%%%%%%%%%%
\begin{figure}[h]
\begin{center}
\includegraphics[trim=25mm 105mm 20mm 50mm,clip,scale=0.45]{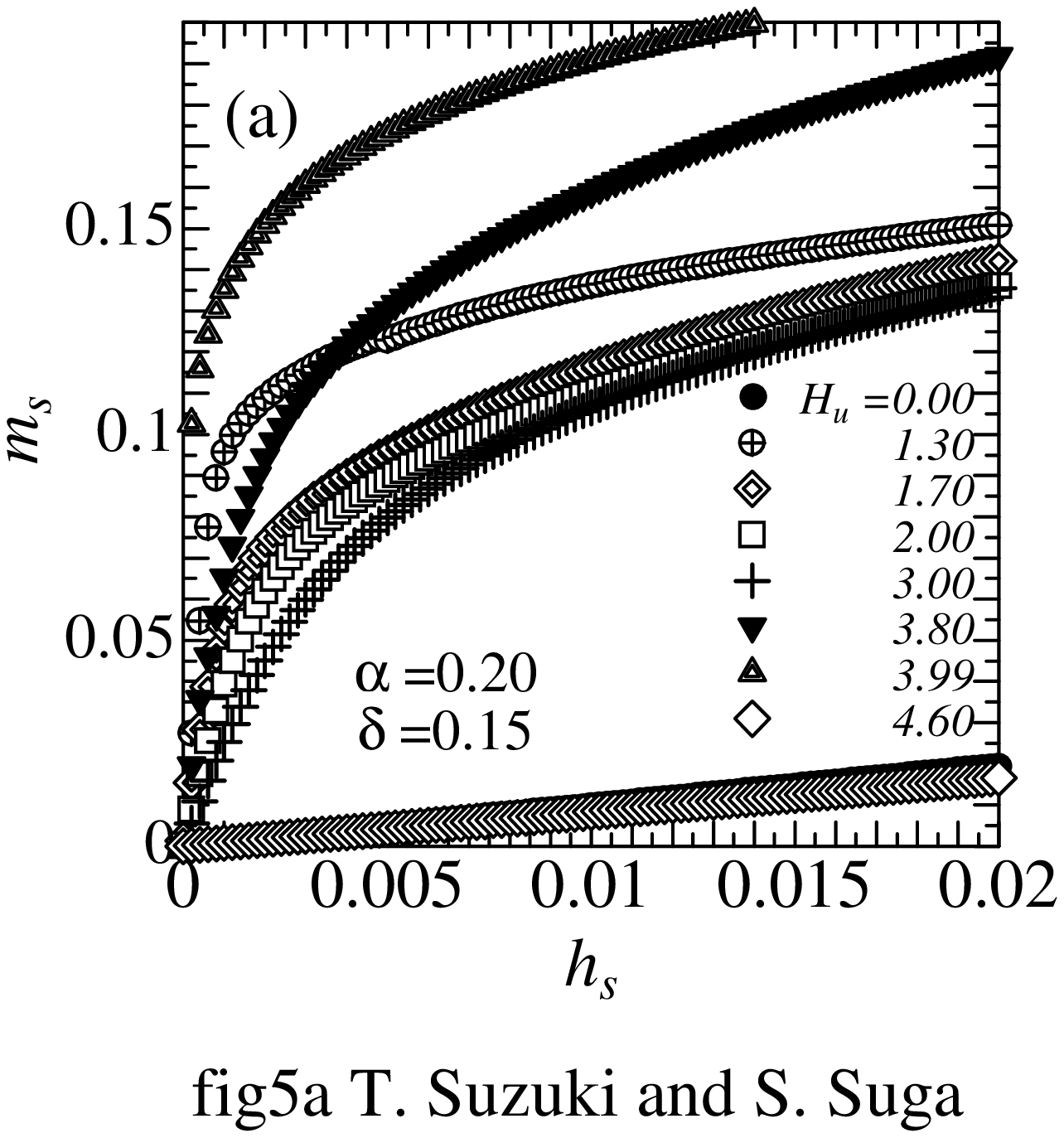}
\includegraphics[trim=25mm 105mm 20mm 50mm,clip,scale=0.45]{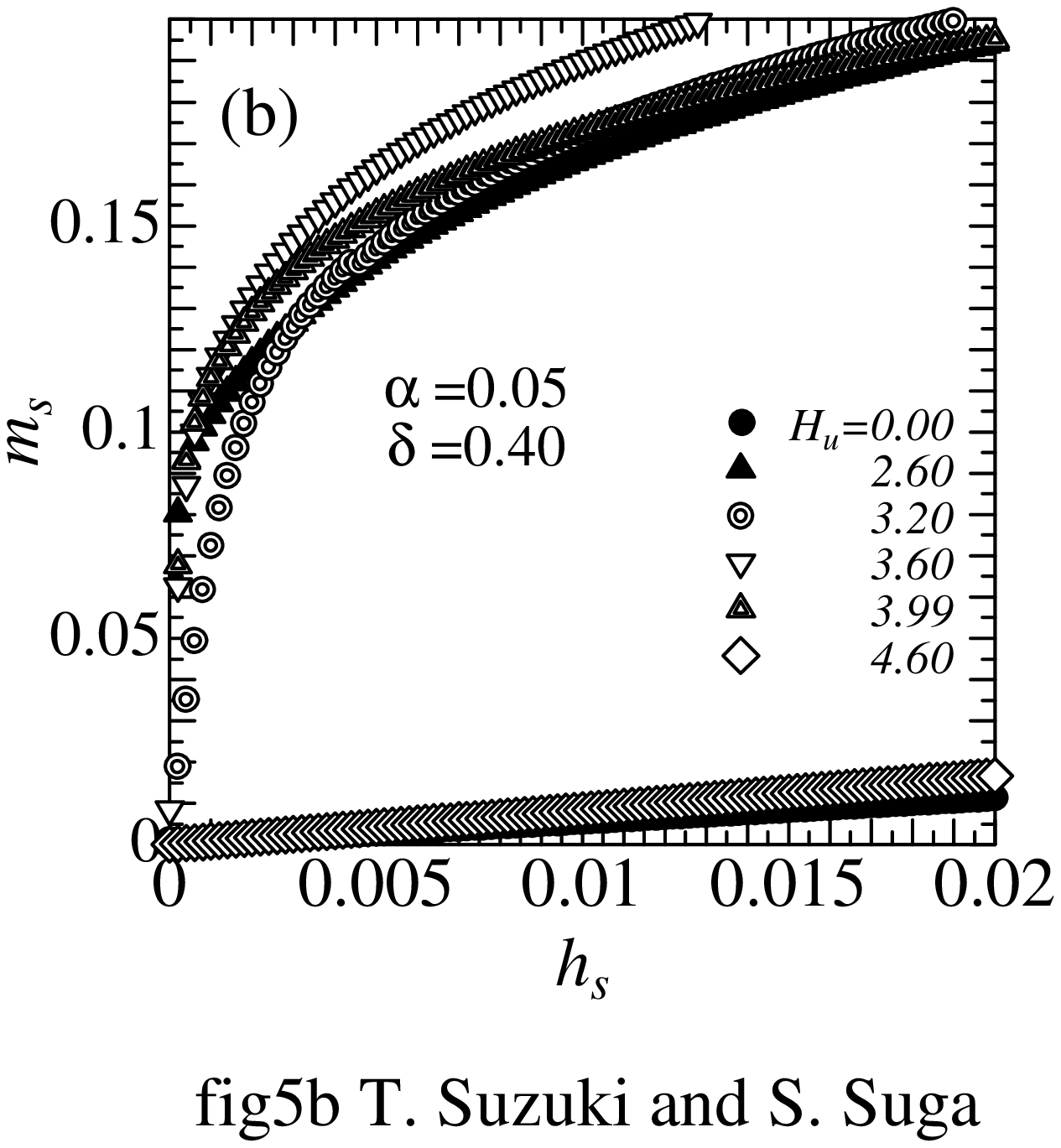}
\end{center}
\vspace{-8mm}
\caption{\small
The staggered magnetization $m_{\rm s}$ for several values of $H_{\rm u}$ as function of $h_{\rm s}$. Parameters used are (a) $\alpha=0.20$ and $\delta=0.15$, and (b) $\alpha=0.05$ and $\delta=0.40$. 
}
\label{fig:fig5}
\end{figure}
%%%%%%%%%%%%%%%%%%%%%%%%%%%%%%%%%%%%%%%%%%%%%%%%%%%%%%%%%%%%%%

We first discuss the results for $\alpha=0.20$ and $\delta=0.15$ shown in Fig. 5(a). At $H_{\rm u}=0$, where the system has an excitation gap, the magnetization curve of $m_{\rm s}$ has a finite derivative at the origin, implying that the staggered susceptibility 
%
%%%%%%%%%%%%%%%%%%%%%%%%%%%%%%%%%%%%%%%%%%%%%%%%%%%%%%%%%%%%%%%%%%%%
\begin{eqnarray}
\chi_{\rm s}= \left.\frac{\partial m_{\rm s}}{\partial h_{\rm s}} \right|_{h_{\rm s} \rightarrow 0} 
\label{eqn:stsus} 
\end{eqnarray}
%%%%%%%%%%%%%%%%%%%%%%%%%%%%%%%%%%%%%%%%%%%%%%%%%%%%%%%%%%%%%%%%%%%%
%
takes a finite value. 
At $H_{\rm u}=4.60$, $m_{\rm u}$ shows the saturation as shown in Fig. 6 and the same behavior of $\chi_{\rm s}$ as in $H_{\rm u}=0$ is observed. 
In $1.30 \leq H_{\rm u} \leq 3.99$, the system becomes gapless. 
At $H_{\rm u}=1.30$ and $3.99$, $\chi_{\rm s}$ shows a divergence, indicating that the staggered long-range order is stabilized. 
In $1.70 \leq H_{\rm u} \leq 3.80$, on the other hand, $\chi_{\rm s}$ takes finite values. The results implies that the staggered long-range order does not emerge up to a certain critical value $zJ^{\prime}_{\rm c}$. 
For $H_{\rm u}=3.00$, the critical value can be evaluated as $zJ^{\prime}_{\rm c} \sim 8.95 \times 10^{-3}$, which is the largest in $1.70 \leq H_{\rm u} \leq 3.80$. The uniform magnetization $m_{\rm u}$ for $\alpha=0.20$ and $\delta=0.15$ in Fig. 6 has been calculated using $zJ^{\prime} = 8.95 \times 10^{-3}$. 
From the results for $m_{\rm u}$, we find that in $1.70 \leq H_{\rm u} \leq 3.80$ the uniform magnetization varies in $0.10 < m_{\rm u} < 0.32$.

In $\alpha=0.05$ and $\delta=0.40$, the behavior of $\chi_{\rm s}$ is quite different as compared with that shown in Fig. 5(a). As shown in Fig. 5(b), $\chi_{\rm s}$ shows a divergence in $2.60<H_{\rm u}<3.99$, where the uniform magnetization varies in $0 < m_{\rm u} \leq 1/2$. The results indicate that the staggered long-range order is always stabilized in $0 \leq m_{\rm u} \leq 1/2$.

%%%%%%%%%%%%%%%%%%%% Fig. 6 %%%%%%%%%%%%%%%%%%%%%%%%%%%%%%%%%%

\begin{figure}[h]
\begin{center}
\includegraphics[trim=25mm 105mm 20mm 40mm,clip,scale=0.45]{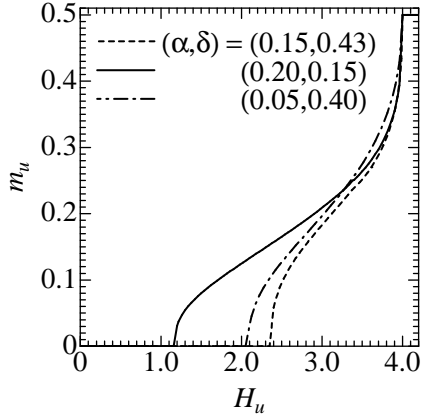}
\end{center}
\vspace{-7mm}
\caption{\small
The uniform magnetization $m_{\rm u}$ for $(\alpha, \delta)=(0.20, 0.15), (0.05, 0.40)$, and $(0.15, 0.43)$ as function of $H_{\rm u}$. The interchain couplings are set to be $zJ^{\prime}=8.95 \times 10^{-3}, 2.50 \times 10^{-2}$, and $6.85 \times 10^{-3}$, respectively.  
}
\label{fig:fig6}
\end{figure}
%%%%%%%%%%%%%%%%%%%%%%%%%%%%%%%%%%%%%%%%%%%%%%%%%%%%%%%%%%%%%%

%%%%%%%%%%%%%%%%%%%%%%%%%%%%%%%%%%%%%%%%%%%%%%%%%%%%%%%%%%%%%%%
\subsection{Static structure factor}
%%%%%%%%%%%%%%%%%%%%%%%%%%%%%%%%%%%%%%%%%%%%%%%%%%%%%%%%%%%%%%%
We investigate characteristics of the ordered state in $1.70 \leq H_{\rm u} \leq 3.80$ for $\alpha=0.20$ and $\delta=0.15$. Since the staggered long-range order does not appear in $zJ<zJ^{\prime}_{\rm c}$ in this case, the staggered magnetization becomes zero and then $h_{\rm s}=0$. 
Using the numerical diagonalization method based on the Lanczos algorithm, we calculate the static structure factor. 
The results are shown in Fig. 7(a).

%%%%%%%%%%%%%%%%%%%% Fig. 7 %%%%%%%%%%%%%%%%%%%%%%%%%%%%%%%%%%
\begin{figure}[h]
\begin{center}
\includegraphics[trim=20mm 150mm 20mm 5mm,clip,scale=0.45]{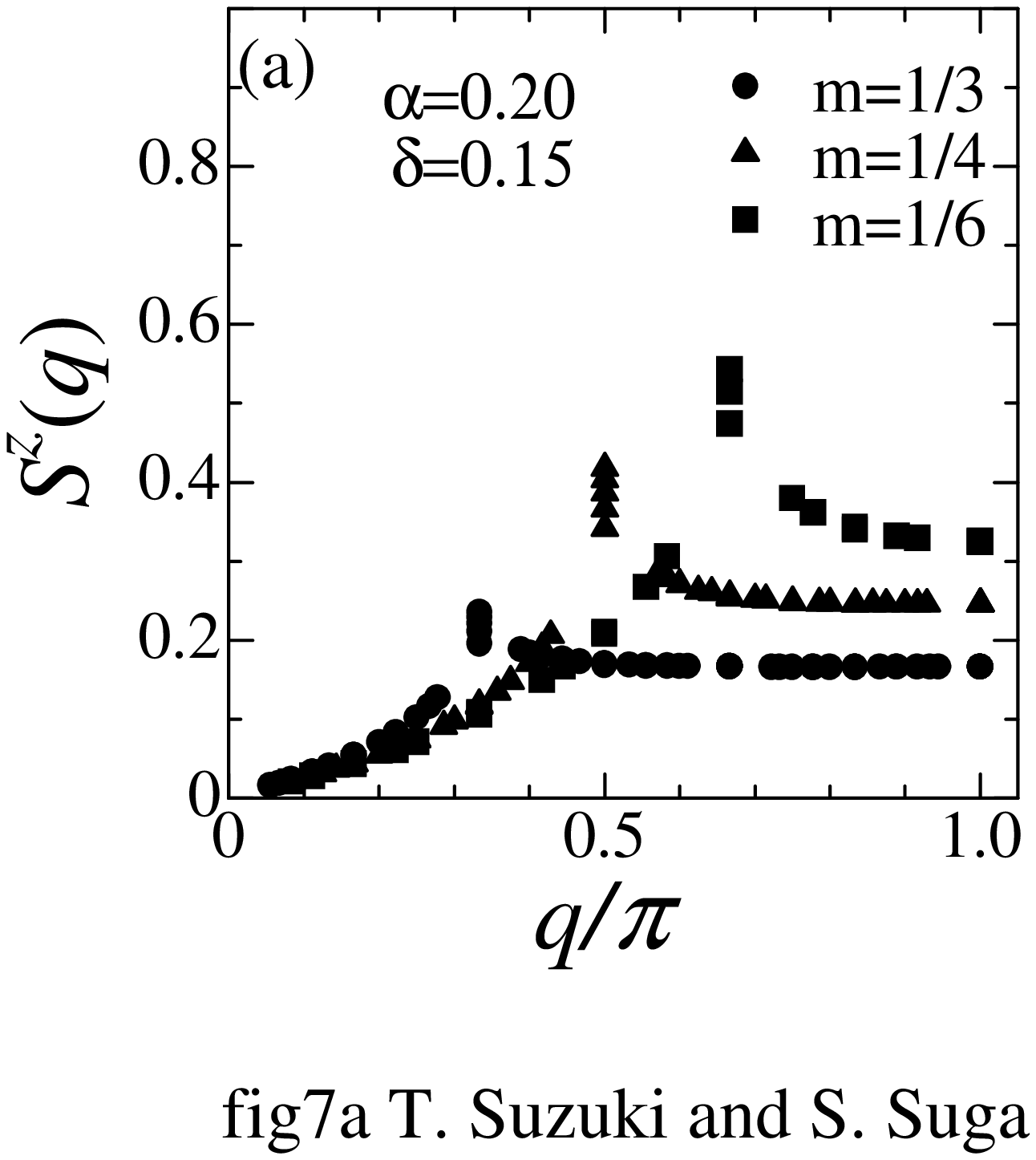}
\includegraphics[trim=20mm 150mm 20mm 5mm,clip,scale=0.45]{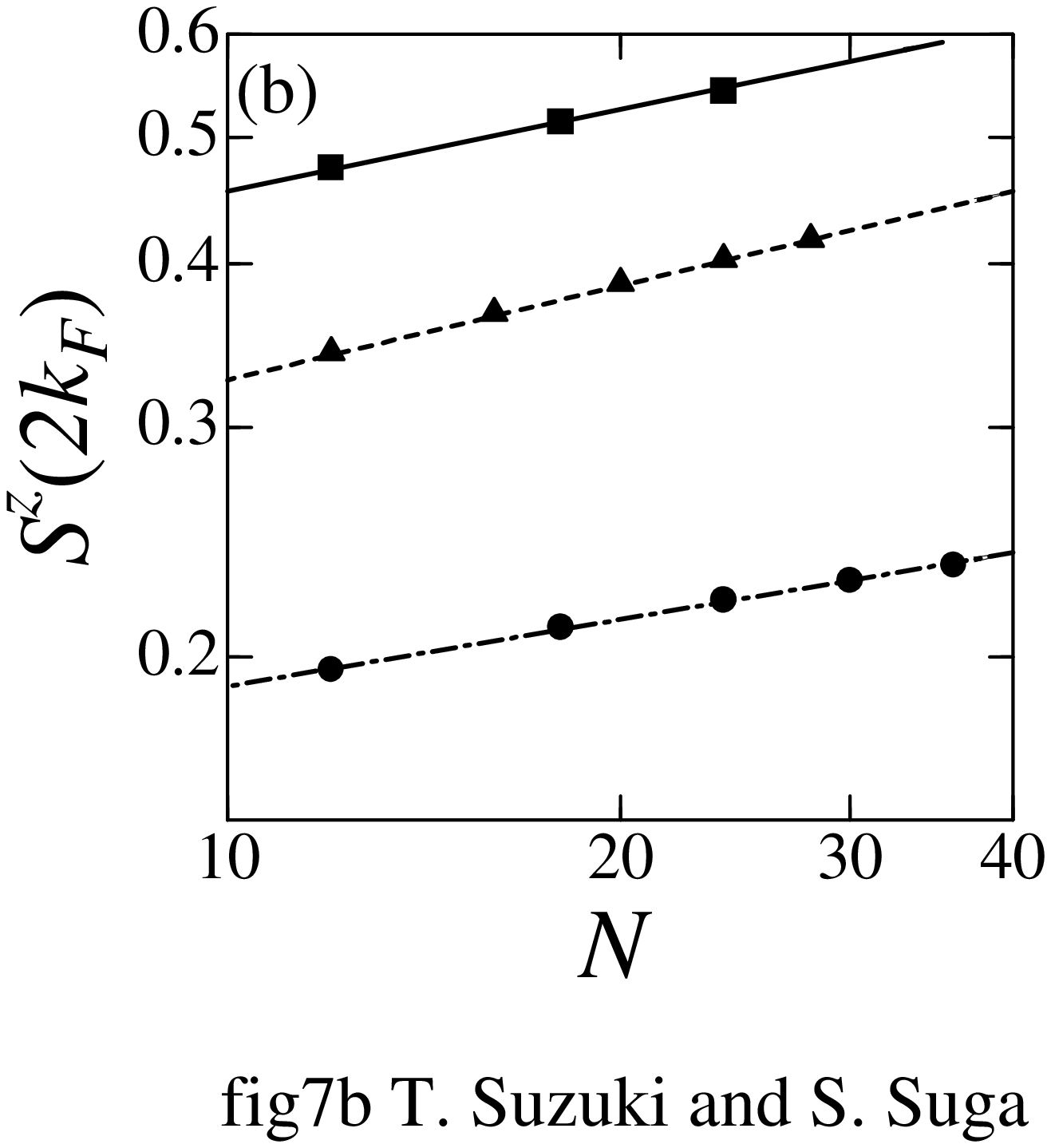}
\end{center}
\vspace{-5mm}
\caption{\small
(a) Static structure factor $S^z(k)$ in $\alpha=0.20$ and $\delta=0.15$ for $m_{\rm u}=1/3, 1/4$, and $1/6$. (b) $N$ dependence of $S^z(2k_{\rm F})$. 
The solid lines are obtained by the least-squares method. The divergence powers are evaluated as $0.196$ for $m_{\rm u}=1/6$, $0.237$ for $m_{\rm u}=1/4$, and $0.170$ for $m_{\rm u}=1/3$. 
}
\label{fig:fig7}
\end{figure}
%%%%%%%%%%%%%%%%%%%%%%%%%%%%%%%%%%%%%%%%%%%%%%%%%%%%%%%%%%%%%%

From the magnetization curve for $m_{\rm u}$ shown in Fig. 6, we estimate that $m_{\rm u}=1/3, 1/4$, and $1/6$ correspond to $H_{\rm u} \sim 3.88, 3.44$, and $2.50$, respectively. For these $H_{\rm u}$, the staggered long-range order does not emerge in $zJ < zJ^{\prime}_{\rm c}$ as shown in Fig. 5(a). 
The size dependence of $S^z(k)$ has an algebraic singularity at $k=2k_{\rm F}$ with increasing the system size $N$ as shown in Fig. 7(b). 
By the least-squares method, the power is evaluated as $0.196$ for $m_{\rm u}=1/6$, $0.237$ for $m_{\rm u}=1/4$, and $0.170$ for $m_{\rm u}=1/3$.  
From the results, we conclude that the system has a tendency towards the formation of an IC long-range order in $1.70 \leq H_{\rm u} \leq 3.80$ for $zJ<zJ^{\prime}_{\rm c}$. 

We calculate $m_{\rm s}$, $m_{\rm u}$, and $S^z(k)$ in other parameter sets of $\alpha$ and $\delta$ systematically using the phase diagram shown in Fig. 2. We find that in the darkly shaded area of Fig. 2 the system has a tendency towards the formation of an IC long-range order with the period $1/m_{\rm u}$ around $m_{\rm u}=1/4$ in $zJ<zJ^{\prime}_{\rm c}$.

%%%%%%%%%%%%%%%%%%%%%%%%%%%%%%%%%%%%%%%%%%%%%%%%%%%%%%%%%%%%%%%%%%%%%%%%%
\section{DISCUSSION}\label{Discussion}
%%%%%%%%%%%%%%%%%%%%%%%%%%%%%%%%%%%%%%%%%%%%%%%%%%%%%%%%%%%%%%%%%%%%%%%%%
\subsection{${\rm F_5PNN}$}
%%%%%%%%%%%%%%%%%%%%%%%%%%%%%%%%%%%%%%%%%%%%%%%%%%%%%%%%%%%%%%%
Recently, $1/T_1$ for ${\rm F_5PNN}$, which is considered to be a 1D $S=1/2$ bond-alternating spin system \cite{hoso}, was measured in magnetic fields \cite{Izumi}. In $H_{c_1}=2.5{\rm T} \leq H \leq H_{c_2}=6.5{\rm T}$, $1/T_1$ exhibited power-law behavior. Furthermore, at $H=5.2{\rm T}$, $1/T_1$ becomes almost independent of temperature. 
It seems difficult to explain the results for $1/T_1$ at $H=5.2{\rm T}$ on the basis of the $S=1/2$ bond-alternating spin-chain model. 
We thus take account of a small NNN interaction in addition to the bond alternation \cite{rhmf}. 

%%%%%%%%%%%%%%%%%%%% Fig. 8 %%%%%%%%%%%%%%%%%%%%%%%%%%%%%%%%%%
\begin{figure}[h]
\begin{center}
\includegraphics[trim=0mm 135mm 30mm 10mm,clip,scale=0.5]{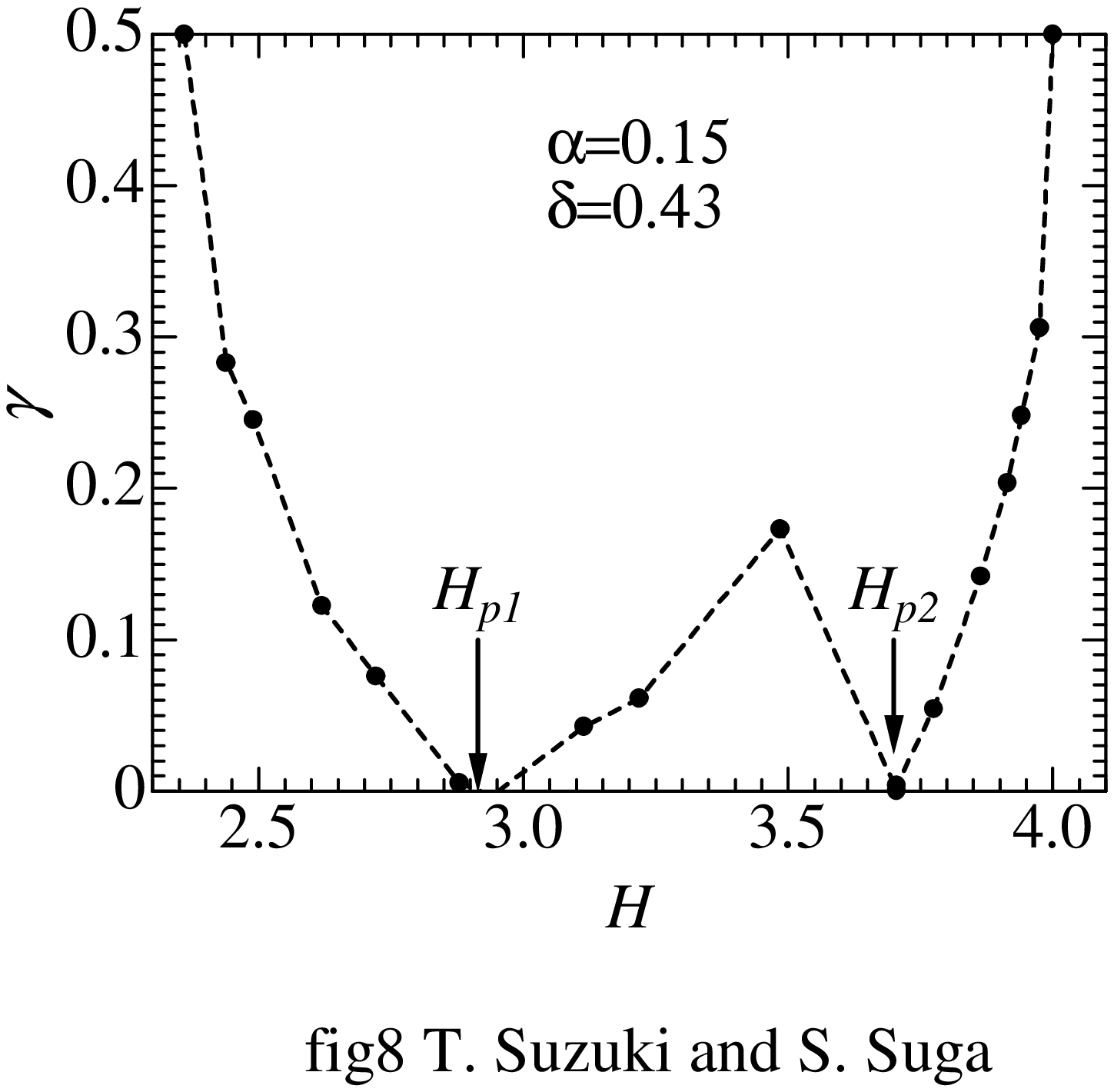}
\end{center}
\vspace{-10mm}
\caption{\small
The divergence exponent $\gamma$ for ${\rm F_5PNN}$ ($\alpha=0.15$ and $\delta=0.43$) as function of $H$, where $1/T_1 \sim T^{-\gamma}$. 
Using the experimental data, $H_{p_1}$ and $H_{p_2}$ are evaluated as 
$H_{p_1} \sim 3.77{\rm T}$ and $H_{p_2} \sim 5.79{\rm T}$. 
The line is to guide the eyes. 
}
\label{fig:fig8}
\end{figure}
%%%%%%%%%%%%%%%%%%%%%%%%%%%%%%%%%%%%%%%%%%%%%%%%%%%%%%%%%%%%%%

On the basis of the Hamiltonian (\ref{eqn:ham1}), we investigate the divergence exponent $\gamma$ of $1/T_1$ for ${\rm F_5PNN}$ numerically. 
Since the bond-alternation parameter of ${\rm F_5PNN}$ was evaluated as $\delta=0.43$ \cite{hoso}, we slightly increase $\alpha$ in a fixed $\delta=0.43$ and investigate whether the behavior $\eta^z<1<\eta^x$ appears in $H_{c_1}<H<H_{c_2}$. As shown in the dotted line in Fig. 2, such behavior emerges at $\alpha \sim 0.15$. The field dependence of $\gamma$ is shown in Fig. 8. 
In $H_{c_1}=2.36<H<H_{c_2}=4.00$ the system is gapless, and at $H_{p_1}=2.88$ and $H_{p_2}=3.71$, $1/T_1$ becomes independent of temperature. 
Using the experimental data of $H_{c_1}$ and $H_{c_2}$, we evaluate the strength of $H_{p_1}$ and $H_{p_2}$;
$H_{p_1} = 2.5{\rm T} + (2.88-2.36)/(4.00-2.36) \times (6.5{\rm T}-2.5{\rm T})
\sim 3.77{\rm T}$ and 
$H_{p_2} = 2.5{\rm T} + (3.71-2.36)/(4.00-2.36) \times (6.5{\rm T}-2.5{\rm T})
\sim 5.79{\rm T}$. 
Therefore, $5.2$T in ${\rm F_5PNN}$ probably corresponds to $H_{p_2}$. 
In order to improve quantitative agreement, a slightly smaller $\alpha$ than $0.15$ may be adequate within $\delta=0.43$. 
We have shown that a small NNN interaction may be essential in ${\rm F_5PNN}$. To confirm this scenario, the magnetic field corresponding to $H_{p_1}$ have to be observed, where $1/T_1$ becomes also independent of temperature. 
If a slightly smaller $\alpha$ than $0.15$ is adequate, $H_{p_1}$ will be slightly larger than $3.77$T. 
Of course, the parameters $\alpha$ and $\delta$ adequate for $1/T_1$ have to well reproduce the experimental results for thermodynamic quantities \cite{hoso,yoshi}. 

%%%%%%%%%%%%%%%%%%%% Fig. 9 %%%%%%%%%%%%%%%%%%%%%%%%%%%%%%%%%%
\begin{figure}[h]
\begin{center}
\includegraphics[trim=5mm 105mm 10mm 50mm,clip,scale=0.45]{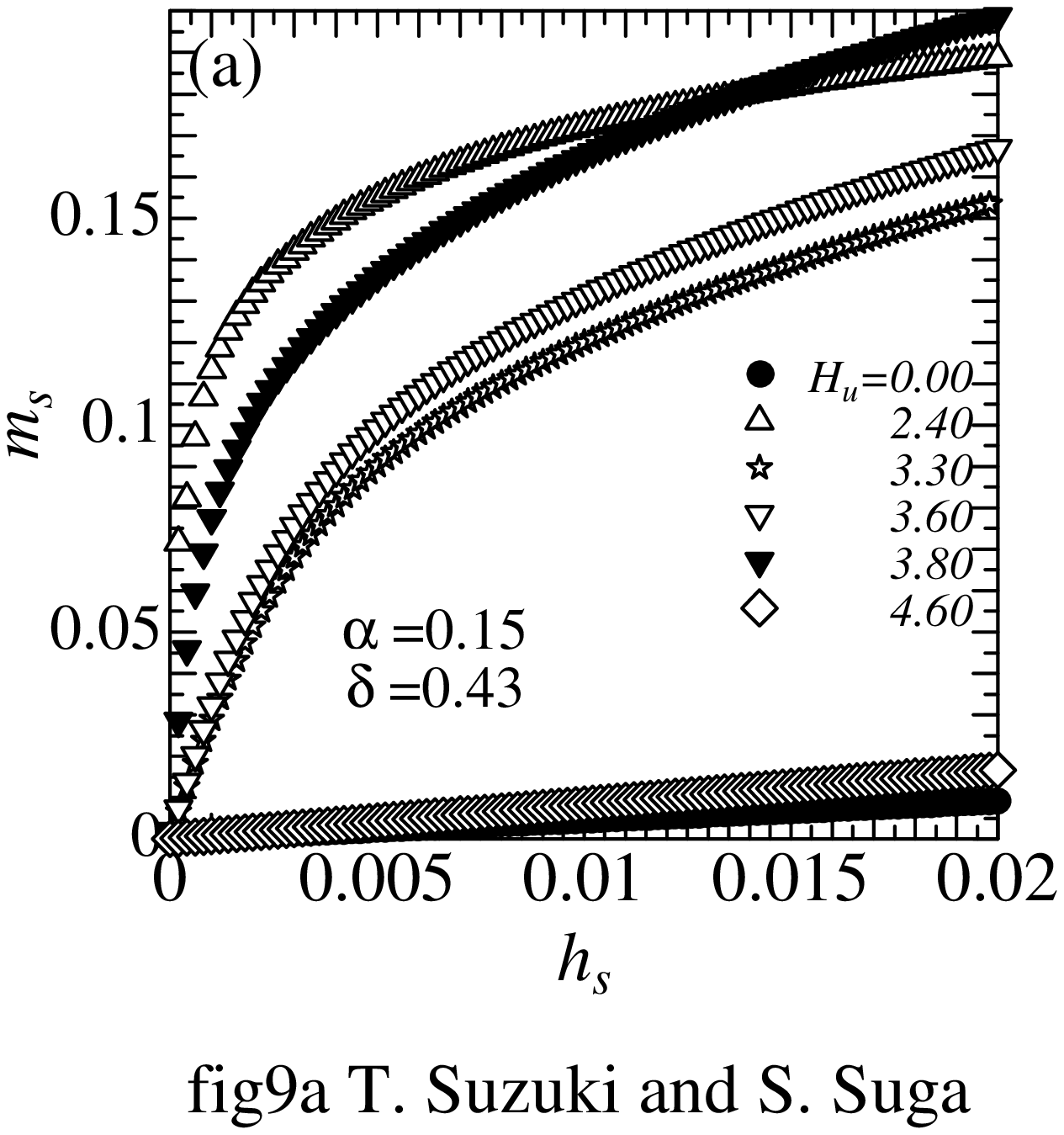}
\includegraphics[trim=5mm 150mm 10mm 5mm,clip,scale=0.45]{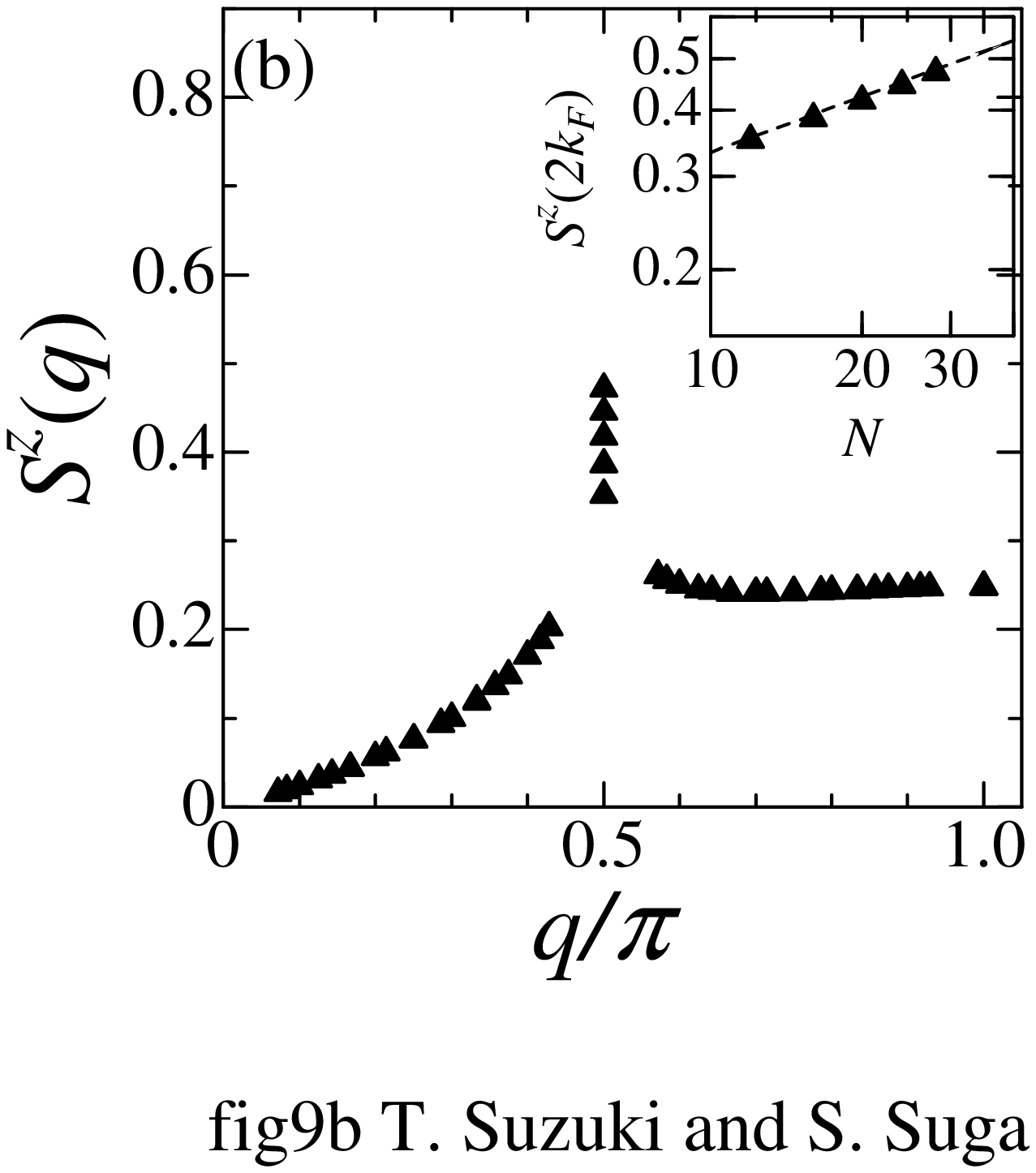}
\end{center}
\vspace{-7mm}
\caption{\small
(a) The staggered magnetization $m_{\rm s}$ for several values of $H_{\rm u}$ as function of $h_{\rm s}$. Parameters are $\alpha=0.15$ and $\delta=0.43$. 
(b) Static structure factor $S^z(k)$ for $m_{\rm u}=1/4$. Inset: $L$ dependence of $S^z(2k_{\rm F})$. From the solid lines obtained by the least-squares method, the divergence power is evaluated as 0.346. 
}
\label{fig:fig9}
\end{figure}
%%%%%%%%%%%%%%%%%%%%%%%%%%%%%%%%%%%%%%%%%%%%%%%%%%%%%%%%%%%%%%%

For ${\rm F_5PNN}$, the transition into the 3D ordered state in magnetic fields was observed in further low temperatures by the specific heat measurements \cite{yoshi}. We discuss the characteristics of the 3D ordered state in ${\rm F_5PNN}$ on the basis of the Hamiltonian (\ref{eqn:ham5}). 
The staggered susceptibility and the static structure factor are calculated for $\alpha=0.15$ and $\delta=0.43$ by the same method as used in Sec. III. 
The results are shown in Fig. 9. 
In $H_{\rm u} = 3.30$ and $3.60$, $\chi_{\rm s}$ takes finite values, implying that the staggered long-range order does not emerge up to a certain critical value $zJ^{\prime}_{\rm c}$.  As shown in Fig. 6, the uniform magnetization takes the values $m_{\rm u}=0.22$ for $H_{\rm u} = 3.30$ and $m_{\rm u}=0.27$ for $H_{\rm u} = 3.60$.  
Note that $m_{\rm u}$ is calculated using the critical value $zJ^{\prime}_{\rm c} = 6.85 \times 10^{-3}$  at $H_{\rm u}=3.60$, which is larger than that at $H_{\rm u}=3.30$. 
In the inset of Fig. 9(a), the size dependence of $S^z(2k_{\rm F})$ for $m_{\rm u} = 1/4 (\sim 0.22$ and $0.27)$ is shown. An algebraic singularity of $S^z(2k_{\rm F})$ as $L \rightarrow \infty$ is seen with the power 0.346. Therefore, the system has a tendency to form an IC long-range order around $m_{\rm u}=1/4$ in $zJ<zJ^{\prime}_{\rm c}$.
 A detailed experimental study for the ordered state of ${\rm F_5PNN}$ under magnetic fields is desirable.

%%%%%%%%%%%%%%%%%%%%%%%%%%%%%%%%%%%%%%%%%%%%%%%%%%%%%%%%%%%%%%%
\subsection{Dominant longitudinal IC spin 
correlations in other systems}
%%%%%%%%%%%%%%%%%%%%%%%%%%%%%%%%%%%%%%%%%%%%%%%%%%%%%%%%%%%%%%%
We have argued that a dominant longitudinal IC spin correlation is attributed to the formation of the half-magnetization plateau, which can be regarded as the CDW insulating state. Therefore, similar behavior is expected in other 1D spin-gapped systems in magnetic fields, {\it e.g.}, the $S=1/2$ two-leg spin-ladder system with a cyclic four-spin interaction. In this system, it was shown that the CDW-like half-magnetization plateau appears \cite{four}.

For a $S=1/2$ two-leg spin-ladder material ${\rm SrCu_2O_3}$, two models were proposed to reproduce the temperature dependence of the susceptibility. 
The one model describes the pure two-leg ladder \cite{Johnson} and the other model includes the effects of a cyclic four-spin interaction \cite{mizuno1}. To distinguish characteristics between the two models, some experimental methods for the observation of dynamical properties were proposed \cite{mizuno2,Haga1,Haga2}. If the temperature-independent $1/T_1$ at certain two fields in $H_{c_1}<H<H_{c_2}$ will be observed, such findings are also an evidence for the effects of a cyclic four-spin interaction in ${\rm SrCu_2O_3}$.

%%%%%%%%%%%%%%%%%%%%%%%%%%%%%%%%%%%%%%%%%%%%%%%%%%%%%%%%%%%%%%%%%%%%%%%%%
\section{SUMMARY}\label{Summary}
%%%%%%%%%%%%%%%%%%%%%%%%%%%%%%%%%%%%%%%%%%%%%%%%%%%%%%%%%%%%%%%%%%%%%%%%%
We have first investigated the critical properties of the $S=1/2$ bond-alternating spin chain with a NNN interaction in magnetic fields. From the results obtained by the numerical calculation and those obtained based on the effective Hamiltonian, we have concluded that there is a parameter region where the longitudinal IC spin correlation becomes dominant around the half-magnetization plateau. 
The results are regarded as a manifestation of the nature of the TL liquid close to the CDW transition point. 
Experimentally, such behavior can be observe by the field dependence of the divergence exponent of $1/T_1$ with decreasing temperature.

When temperature is further decreased, the interchain interaction becomes relevant. We have calculated the staggered susceptibility and the uniform magnetization, combining the DMRG method with the interchain mean-field theory. 
In the parameter region where the dominant longitudinal IC spin correlation appears for $J^{\prime}=0$, the staggered long-range order does not emerge up to a certain critical value $zJ^{\prime}_{\rm c}$ around $m_{\rm u} = 1/4$, while in the other parameter region, the staggered long-range order is stabilized in $0 \leq m_{\rm u} \leq 1/2$. 
To investigate the characteristics of the long-range order in the former parameter region, we have calculated the static structure factor. 
From the size dependence of $S^z(2k_{\rm F})$, we have shown that the system has a tendency to form an IC long-range order around $m_{\rm u}=1/4$ in $zJ<zJ^{\prime}_{\rm c}$. 
Using the results, we have discussed the recent experimental results for $1/T_1$ in magnetic fields performed for ${\rm F_5PNN}$.

%%%%%%%%%%%%%%%%%%%%%%%%%%%%%%%%%%%%%%%%%%%%%%%%%%%%%%%%%%%%%%%%%%%%%%%%%
\section{Acknowledgments}
%%%%%%%%%%%%%%%%%%%%%%%%%%%%%%%%%%%%%%%%%%%%%%%%%%%%%%%%%%%%%%%%%%%%%%%%%
We would like to thank T. Goto (Kyoto University), T. Goto (University of Tokyo), Y. Hosokoshi, K. Izumi, A. Kawaguchi, N. Maeshima, S. Miyashita, T. Sakai, and Y. Yoshida for useful comments and valuable discussions. 
Part of our computational programs are based on TITPACK version 2 by H. Nishimori. Numerical computations were carried out at the Yukawa Institute Computer Facility, Kyoto University, and the Supercomputer Center, the Institute for Solid State Physics, University of Tokyo. 
This work was supported by a Grant-in-Aid for Scientific Research from the Ministry of Education, Culture, Sports, Science, and Technology, Japan.

%%%%%%%%%%%%%%%%%%%%%%%%%%%%%%%%%%%%%%%%%%%%%%%%%%%%%%%%%%%%%%%%%%%%%%%%%%%%%
%                            REFERENCES                                     %
%%%%%%%%%%%%%%%%%%%%%%%%%%%%%%%%%%%%%%%%%%%%%%%%%%%%%%%%%%%%%%%%%%%%%%%%%%%%%
% Create the reference section using BibTeX
%\bibliography{nmr}

\begin{references}
\bibitem{comm}
In the darkly shaded area of Fig. 2, we have confirmed that $H^+(1/4)$ and $H^-(1/4)$ in (\ref{eqn:as1}) and (\ref{eqn:as2}) extrapolated into $L \rightarrow \infty$ satisfy $H^+(1/4)=H^-(1/4)$ within the accuracy of at least $O(10^{-3})$.
\end{references}
%%%%%%%%%%%%%%%%%%%%%%%%%%%%%%%%%%%%%%%%%%%%%%%%%%%%%%%%%%%%%%%%%%%%%%%%%%%%%

%

%

\end{document}